\RequirePackage[dvipsnames,table,prologue]{xcolor} 
\RequirePackage[T1]{fontenc}
\RequirePackage{lmodern}

\documentclass[runningheads]{llncs}

\RequirePackage{filecontents}
\PassOptionsToPackage{dvipsnames}{xcolor}
\RequirePackage{comgipp}
\RequirePackage{amsfonts}
\RequirePackage{amsmath}
\RequirePackage{authblk}

\usepackage[%
	paperheight=235mm,
	paperwidth=155mm,
	textwidth=12.2cm,
	textheight=19.3cm,
	includehead]
{geometry}
\usepackage{bbding}

\makeatletter
\RequirePackage[bookmarks,unicode,colorlinks=true,breaklinks]{hyperref}%
   \def\@citecolor{blue}%
   \def\@urlcolor{blue}%
   \def\@linkcolor{blue}%

\def\orcidID#1{\smash{\href{http://orcid.org/#1}{\protect\raisebox{-1.25pt}{\protect\includegraphics{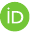}}}}}
\makeatother

\usepackage{amssymb}
\usepackage{amsmath}

\usepackage[
	backend=biber,
	sortcites,
	safeinputenc
]{biblatex}

\makeatletter
\newcommand*{\LaCASt}{{L\kern -.36em{\sbox \z@ T\vbox to\ht \z@ {\hbox {\check@mathfonts \fontsize \sf@size \z@ \math@fontsfalse \selectfont A}\vss }}\kern -.05em C\kern-.2em{\sbox\z@ T\vbox to\ht\z@{\hbox{\check@mathfonts\fontsize\sf@size\z@\math@fontsfalse\selectfont\scriptsize \lower.89em\hbox{AS}}\vss}}\kern-.15em\relax{\scshape T}}}

\makeatother

\usepackage{graphicx} 
\graphicspath{{images/}} 
\usepackage{caption}
\usepackage{subcaption}
\usepackage{wrapfig}
\usepackage{mathrsfs}
\usepackage{latexml}
\usepackage{soul}
\usepackage{array}
\usepackage{makecell}
\usepackage{cellspace}
\usepackage{arydshln}
\usepackage{multirow}
\usepackage{verbdef}

\usepackage[most]{tcolorbox}

\tcbset{colback=Gray!3!white, colframe=Blue!50!white, left=0mm, right=0mm, top=0mm, bottom=0mm,
        highlight math style= {enhanced, 
            colframe=red,colback=red!10!white,boxsep=0pt}
        }

\usepackage{enumitem} 
\usepackage{calc}
\usepackage[skip=1.5\baselineskip]{caption}
\usepackage{stackengine}
\usepackage{cutwin}

\SetEnumitemKey{midsep}{topsep=2.5pt, partopsep=2.5pt}

\newcommand{\NN}{{\mathbb{N}}}

\newcommand{\RR}{{\mathbb{R}}}
\newcommand{\CC}{{\mathbb{C}}}

\newcommand\Maple[1]{{\tt Maple}#1}
\newcommand\Mathematica[1]{{\tt Mathematica}#1}
\newcommand\Map[1]{{\tt Maple}#1}
\newcommand\Math[1]{{\tt Mathematica}#1}

\newlength\myheight
\newlength\mydepth
\settototalheight\myheight{Xygp}
\newcommand*\inlinegraphics[1]{%
  \settototalheight\myheight{Xygp}%
  \settodepth\mydepth{Xygp}%
  \raisebox{-\mydepth}{\includegraphics[height=\myheight]{#1}}%
}

\newcolumntype{R}[1]{>{\raggedleft\let\newline\\\arraybackslash\hspace{0pt}}m{#1}}

\usepackage{tikz}
\usetikzlibrary{calc}

\usepackage{stfloats}

\usepackage{censor}
\usepackage[subtle]{savetrees}

\usepackage{pifont}
\newcommand{\correct}{\textcolor{OliveGreen}{\textbf{\ding{51}}}}

\newcommand{\wrong}{\textcolor{Red!20!BrickRed}{\textbf{\ding{56}}}}

\DeclareBibliographyCategory{preprintref}
\newcommand{\setPreprintRef}[1]{%
\def\theReferenceText{\fullcite{#1}}\addtocategory{preprintref}{#1}}

\setPreprintRef{GreinerPetter2022}
\newcommand{\dlmf}[1]{\cite[\href{https://dlmf.nist.gov/#1}{(#1)}]{DLMF}}

\colorlet{TableRowColor}{Gray!20!white}
\colorlet{CorrectBackgroundDLMFColor}{SpringGreen!90}

\newcommand\demolink{\url{https://lacast.wmflabs.org}}
\newcommand{\lacastLink}[2]{\href{https://lacast.wmflabs.org/wiki/#1}{\texttt{#2}}}

\input{codeboxes.tex}

\begin{document}

\title{Comparative~Verification of the Digital Library of Mathematical Functions and Computer Algebra Systems}
\titlerunning{Comparative Verification of the DLMF and CAS}
\author{%
	Andr\'{e} Greiner-Petter\inst{1}(\Envelope)\orcidID{0000-0002-5828-5497} \and
	Howard~S.~Cohl\inst{2}\orcidID{0000-0002-9398-455X} \and
	Abdou~Youssef\,\inst{2,3} \and
	Moritz Schubotz\inst{1,4}\orcidID{0000-0001-7141-4997} \and %
	Avi Trost\inst{5} \and
	Rajen Dey\inst{6} \and
	Akiko Aizawa\inst{7}\orcidID{0000-0001-6544-5076} \and %
	Bela Gipp\inst{1}\orcidID{0000-0001-6522-3019} %
}

\institute{
	University of Wuppertal, Wuppertal, Germany,\\%
	\email{\{greinerpetter,schubotz,gipp\}@uni-wuppertal.de}
\and
	National Institute of Standards and Technology,\\ Mission Viejo, CA, U.S.A., %
	\email{howard.cohl@nist.gov}
\and
	George Washington University, Washington, D.C., U.S.A, %
	\email{ayoussef@gwu.edu}
\and
	FIZ Karlsruhe, Berlin, Germany, %
	\email{moritz.schubotz@fiz-karlsruhe.de}
\and
	Brown University, Providence, RI, U.S.A., %
	\email{avitrost@gmail.com}
\and
	University of California Berkeley, Berkeley, CA, U.S.A., %
	\email{rajhataj@gmail.com}
\and
	National Institute of Informatics, Tokyo, Japan, %
	\email{aizawa@nii.ac.jp}
}

\authorrunning{A.~Greiner-Petter et al.}

\maketitle
\thispagestyle{firststyle}

\vspace*{-0.4cm}
\begin{abstract}
Digital mathematical libraries assemble the knowledge of years 
of mathematical research. Numerous disciplines (e.g., physics, 
engineering, pure and applied mathematics) rely heavily on 
compendia gathered findings.
Likewise, modern research applications rely more and more on 
computational solutions, which are often calculated and verified 
by computer algebra systems. Hence, the correctness, accuracy, 
and reliability of both digital mathematical libraries and computer 
algebra systems is a crucial attribute for modern research.
In this paper, we present a novel approach to verify a digital 
mathematical library and two computer 
algebra systems with one another by converting mathematical 
expressions from one system to the other.
We use our previously developed conversion tool (referred to as \LaCASt) 
to translate formulae from the NIST Digital Library of 
Mathematical Functions to the computer algebra systems \Maple{}
and \Mathematica.
The contributions of our presented work are as follows:~(1) we 
present the most comprehensive verification of computer algebra 
systems and digital mathematical libraries with one another;
(2) we significantly enhance the performance of the underlying 
translator in terms of coverage and accuracy; and
(3) we provide open access to translations for \Maple{} and
\Mathematica{} of the formulae in the NIST Digital Library of 
Mathematical Functions.


\end{abstract}

\vfill
\vspace*{-0.4cm}
\section{Introduction}\label{sec:introduction}
Digital Mathematical Libraries (DML) gather the knowledge and results 
from thousands of years of mathematical research.
Even though pure and applied mathematics are precise disciplines, 
gathering their knowledge bases over 
many years results in issues 
which every digital library shares:
~consistency, completeness, and accuracy. 
Likewise, Computer Algebra Systems (CAS)\footnote{In the sequel, the 
acronyms CAS and DML are used, depending on the context,
interchangeably with their plurals.} play a crucial role in the modern 
era for pure and applied mathematics, 
and those fields which rely on them.
CAS can be used to simplify, 
manipulate, compute, and visualize mathematical 
expressions. 
Accordingly, modern research regularly uses DML and CAS together.
Nonetheless, DML~\cite{Cohl18,Greiner-Petter19} and 
CAS~\cite{CarefulCAS,kaliszyk2007certified,Duran2014} are not 
exempt from having 
bugs or errors. Dur\'{a}n et al.~\cite{Duran2014} even raised the 
rather dramatic question:~``\textit{\!can we trust in [CAS]}?''

Existing comprehensive DML, such as the Digital Library of Mathematical Functions (DLMF)~\cite{DLMF}, 
are consistently updated and frequently corrected with errata%
\footnote{\url{https://dlmf.nist.gov/errata/} [accessed 09/01/2021]}.
Although each chapter of the DLMF 
has been carefully written, edited, validated, and proofread over many years, 
errors still remain.
Maintaining a DML, such as the DLMF, is a laborious process.
Likewise, CAS are eminently complex systems, and in the case of commercial 
products, often similar to black boxes in which the 
magic (i.e., the computations) happens in opaque private code~\cite{Duran2014}.
CAS, especially commercial products, are often exclusively tested internally during development.

An independent examination process can improve testing and increase trust 
in the systems and libraries.
Hence, we want to elaborate on the following research question.
\vspace*{-0.05cm}
\begin{tcolorbox}[width=\linewidth, colframe=MidnightBlue!40!black, colback=Gray!12, boxsep=0.20cm, arc=1mm, boxrule=0.2mm, leftrule=1mm]
How can digital mathematical libraries and computer algebra systems be 
utilized to improve and verify one another?
\end{tcolorbox}
\vspace*{-0.1cm}
Our initial approach for answering this question is inspired by our previous studies on translating DLMF equations to CAS~\cite{Cohl18}. 
In order to verify a translation tool from a specific \LaTeX{} dialect to \Maple
\footnote{%
The mention of specific products, trademarks, or brand names is for purposes of 
identification only. Such mention is not to be interpreted in any way as an endorsement
or certification of such products or brands by the National Institute of Standards and
Technology, nor does it imply that the products so identified are necessarily the best
available for the purpose. All trademarks mentioned herein belong to their respective
owners.%
}.
, we performed \textit{symbolic} and \textit{numeric} evaluations on equations from the DLMF. Our 
approach presumes that a proven equation in a DML must be also valid in a CAS. In 
turn, a disparity in between the DML and CAS would lead to an issue in the 
translation process. 
However, assuming a correct translation, a disparity would 
also indicate an issue either in the DML source or the CAS implementation. In turn, 
we can take advantage of the same approach to 
improve and even verify DML with CAS and vice versa.
Unfortunately, previous efforts to translate mathematical expressions from various 
formats, such as \LaTeX~\cite{Cohl17,Greiner-Petter19,Parisse17}, \MathML~\cite{SchubotzGSMCG18}, 
or OpenMath~\cite{HerasPR09,PrietoDP00}, to CAS syntax have shown that the translation will be the most critical part of this verification approach.

In this paper, we elaborate on the feasibility and limitations of the translation 
approach from DML to CAS as a possible answer to our research question. We further 
focus on the DLMF as our DML and the two general-purpose CAS \Maple{} and \Mathematica{} 
for this first study. This relatively sharp limitation is necessary in order to analyze 
the capabilities of the underlying approach to verify commercial CAS and large DML. The 
DLMF uses semantic macros internally in order to disambiguate mathematical expressions
~\cite{MillerY03,YoussefM20}. These macros help to mitigate the open issue of 
retrieving sufficient semantic information from a context to perform translations to 
formal languages~\cite{SchubotzGSMCG18,Greiner-Petter19}. Further, the DLMF and 
general-purpose CAS have a relatively large overlap in coverage of special functions and 
orthogonal polynomials. Since many of those functions play a crucial role in a large 
variety of different research fields, we focus in this study mainly on these 
functions. 
Lastly, we will take our previously developed translation tool \LaCASt~\cite{Cohl17,Greiner-Petter19} as the baseline for translations from the DLMF to \Maple.
In this successor project, we focus on improving \LaCASt{} to 
minimize the negative effect of wrong translations as much as possible for our study. 
In the future, other DML and CAS can be improved and verified following the same 
approach by using a different translation approach depending on the data of the DML, 
e.g., \MathML~\cite{SchubotzGSMCG18} or OpenMath~\cite{HerasPR09}.

In particular, in this paper, we fix the majority of the remaining issues of \LaCASt~\cite{Cohl18}, which allows our tool to translate twice as many expressions 
from the DLMF to the CAS as before.
Current extensions include the support for the mathematical operators:~sum, product, limit, 
and integral, as well as overcoming semantic hurdles associated with Lagrange (prime) notations 
commonly used for differentiation.
Further, we extend its support to 
include \Mathematica{} using the freely available \emph{Wolfram Engine for Developers} 
(WED)\footnote{\url{https://www.wolfram.com/engine/} [accessed 09/01/2021]} (hereafter, with \Mathematica, we refer to the WED).
These improvements allow us to cover a larger portion of the DLMF, increase the reliability
of the translations via \LaCASt, and allow for comparisons between two major
general-purpose CAS for the first time, namely \Maple{} and \Mathematica.
Finally, we provide open access to all the results contained within this paper, including all translations of DLMF formulae, an endpoint to \LaCASt\footnote{\label{ft:data}\url{https://lacast.wmflabs.org/} [accessed 01/01/2022]}, and the full source code of \LaCASt\footnote{\url{https://github.com/ag-gipp/LaCASt} [accessed 04/01/2022]}.

The paper is structured as follows. Section~\ref{sec:dlmf} explains the data in the DLMF. Section~\ref{sec:lacast} focus on the improvements of \LaCASt{} that had been made to make the translation as comprehensive and reliable as possible for the upcoming evaluation. Section~\ref{sec:evaluation} explains the symbolic and numeric evaluation pipeline. Since Cohl et al.~\cite{Cohl18} only briefly sketched the approach of a numeric evaluation, we will provide an in-depth discussion of that process in Section~\ref{sec:evaluation}. Subsequently, we analyze the results in Section~\ref{sec:results}. Finally, we conclude the findings and provide an outlook for upcoming projects in Section~\ref{sec:future-work}.

\subsection{Related Work}
Existing verification techniques for CAS often focus on specific subroutines or functions~\cite{CompareCAS,kaliszyk2007certified,CertifyCAS1,Elphick2003,Carette2011,%
Leino2012,Khan2014,HarrisonT98}, such as a specific theorems~\cite{Lamban2013}, differential equations~\cite{HickmanLF21}, or the implementation of the \verb|math.h| library~\cite{Lee2018}.
Most common are verification approaches that rely on intermediate verification languages~\cite{CertifyCAS1,kaliszyk2007certified,Khan2014,HickmanLF21,HarrisonT98}, 
such as \textit{Boogie}~\cite{Leino2012,Boogie2006} or \textit{Why3}~\cite{Khan2014,Why32011},
which, in turn, rely on proof assistants and theorem provers, such as \textit{Coq}~\cite{CertifyCAS1,Bertot2004}, \textit{Isabelle}~\cite{HickmanLF21,Nipkow2002}, or \textit{HOL Light}~\cite{kaliszyk2007certified,Harrison96,HarrisonT98}.
Kaliszyk and Wiedijk~\cite{kaliszyk2007certified} proposed on entire new CAS which is built on top of the proof assistant HOL Light so that each simplification step can be proven by the underlying architecture.
Lewis and Wester~\cite{CompareCAS} manually compared the symbolic computations on polynomials and matrices with seven CAS.
Aguirregabiria et al.~\cite{CarefulCAS} suggested to teach students the known traps and difficulties with evaluations in CAS instead to reduce the overreliance on computational solutions.

Cohl et al.~\cite{Cohl18} developed the aforementioned translation tool \LaCASt, which translates expressions from a semantically enhanced \LaTeX{} dialect to \Maple. By evaluating the performance and accuracy of the translations, we were able to discover a sign-error in one the DLMF's equations~\cite{Cohl18}. While the evaluation was not intended to verify the DLMF, the translations by the rule-based translator \LaCASt{} provided sufficient robustness to identify issues in the underlying library. 
To the best of our knowledge, besides this related evaluation via \LaCASt, there are no existing libraries or tools that 
allow for automatic verification of DML.

\section{The DLMF dataset}\label{sec:dlmf}
In the modern era, most mathematical texts (handbooks, journal publications, magazines, monographs, treatises, 
proceedings, etc.) are written using the document preparation system \LaTeX. 
However, the focus of \LaTeX{} is for precise control of the rendering mechanics 
rather than for a semantic description of its content.
In contrast,
CAS syntax is coercively unambiguous in order to
interpret the input correctly. Hence, a transformation tool from DML to CAS must
disambiguate mathematical expressions. While there is an ongoing effort towards
such a process~\cite{Schubotz2016,Kristianto2017,Youssef17,GreinerPetter2020,ARQMath,ShanY21}, there is
no reliable tool available to disambiguate mathematics sufficiently to date.

The DLMF contains numerous relations between functions and many other properties. 
It is written in \LaTeX{} but uses specific semantic macros when applicable~\cite{YoussefM20}.
These semantic macros represent a unique function or polynomial defined in the DLMF.
Hence, the semantic \LaTeX{} used in the DLMF is often unambiguous.
For a successful evaluation via CAS, we also need to
utilize all requirements of an equation, such as constraints, domains, or substitutions.
The DLMF provides this additional data too and generally in a machine-readable form~\cite{YoussefM20}.
This data is accessible via the i-boxes (information boxes next to an equation marked with the icon {\Large \inlinegraphics{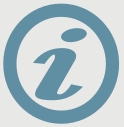}}).
If the information is not given in the attached i-box or the information is incorrect, the translation via \LaCASt{} would fail.
The i-boxes, however, do not contain information about branch cuts (see 
Section~B) 
or constraints.
Constraints are accessible if they are directly attached to an equation. If they appear in the text (or even a title), \LaCASt{} cannot utilize them.
The test dataset, we are using, was generated from DLMF Version 1.1.3 (2021-09-15) and
contained $9,\!977$ formulae with $1,\!505$ defined symbols, $50,\!590$ used symbols, $2,\!691$ constraints, and $2,\!443$ warnings for non-semantic expressions, i.e., expressions without semantic macros~\cite{YoussefM20}. Note that the DLMF does not provide access to the underlying \LaTeX{} source. Therefore, we added the source of every equation to our result dataset.

\vspace{-0.1cm}
\section{Semantic \LaTeX{} to CAS translation}\label{sec:lacast}
The aforementioned translator \LaCASt{} was developed by Cohl and 
Greiner-Petter et al.~\cite{Cohl17,Cohl18,Greiner-Petter19}.
They reported a coverage of 58.8\% translations for a manually selected
part of the DLMF to the CAS \Maple. This version of \LaCASt{} serves
as a baseline for our improvements. In order to verify their translations,
they used symbolic and numeric evaluations and reported a success rate of
$\sim\!\!16\%$ for symbolic and $\sim\!\!12\%$ for numeric verifications.

Evaluating the baseline on the entire DLMF result in a coverage of only 
31.6\%. Hence, we first want to increase the coverage of \LaCASt{} on the DLMF.
To achieve this goal, we first increasing the number of translatable semantic macros
by manually defining more translation patterns for special functions and orthogonal polynomials.
For \Maple, we increased the number from 201 to 261. For \Mathematica, 
we define 279 new translation patterns which enables \LaCASt{} to perform 
translations to \Mathematica. Even though the DLMF uses 675 distinguished 
semantic macros, we cover $\sim\!\!70\%$ of all DLMF equations with our
extended list of translation patterns (see Zipf's law for mathematical 
notations~\cite{GreinerPetter2020a}).
In addition, we implemented rules for translations that are applicable in the context 
of the DLMF, e.g., ignore ellipsis following floating-point values or $\verb|\choose|$ 
always refers to a binomial expression.
Finally, we tackle the remaining issues outlined by Cohl et al.~\cite{Cohl18} which
can be categorized into three groups:~(i) expressions of which the arguments of 
operators are not clear, namely sums, products, integrals, and limits; (ii) 
expressions with prime symbols indicating differentiation; and (iii) expressions 
that contain ellipsis.
While we solve some of the cases in Group (iii) by ignoring ellipsis following floating-point values, most of these cases remain unresolved. In the following, we elaborate our solutions
for (i) in Section~\ref{subsec:limited-expression} and (ii) in Section~\ref{subsec:lagrange-notation}.

\vspace{-0.3cm}
\subsection{Parse sums, products, integrals, and limits}
\label{subsec:limited-expression}
Here we consider common notations for the sum, product, integral, and limit operators. For these operators, one may
consider mathematically essential operator metadata (MEOM). For all these operators, the 
MEOM includes {\it argument(s)} and {\it bound variable(s)}.  The operators act on the arguments,
which are themselves functions of the bound variable(s). For sums and products, the bound variables 
are referred to as \emph{indices}. The bound variables for integrals\footnote{The notion of integrals includes:~antiderivatives (indefinite integrals), 
definite integrals, contour integrals, multiple (surface, volume, etc.) integrals, 
Riemannian volume integrals, Riemann integrals, Lebesgue integrals, Cauchy principal value 
integrals, etc.} are called \emph{integration variables}. For limits, the bound variables 
are continuous variables (for limits of continuous functions) and indices (for limits of sequences).
For integrals, MEOM include precise descriptions of regions of integration
(e.g., piecewise continuous paths/intervals/regions). For limits, MEOM include
limit points (e.g., points in $\RR^n$ or $\CC^n$ for $n \!\in\! \NN$), as well as information
related to whether the limit to the limit point is independent or dependent on the direction in which 
the limit is taken (e.g., one-sided limits).

For a translation of mathematical expressions involving the \LaTeX{}
commands \verb|\sum|, \verb|\int|, \verb|\prod|, and \verb|\lim|, we must extract
the MEOM.  This is achieved by
(a) determining the argument of the operator and 
(b) parsing corresponding subscripts, superscripts, and arguments.
For integrals, the MEOM may be complicated, but certainly contains the argument
(function which will be integrated), bound (integration) variable(s) and details
related to the region of integration.
Bound variable extraction is usually straightforward
since it is usually contained within a differential 
expression (infinitesimal, pushforward, differential 1-form, exterior derivative, 
measure, etc.), e.g., $\mathrm{d}x$. 
Argument extraction is less straightforward since even though differential expressions 
are often given at the end of the argument, sometimes the differential expression
appears in the numerator of a fraction (e.g., $\int\!\tfrac{f(x)\mathrm{d}x}{g(x)}$).
In which case, the argument is everything to the right of the
\verb|\int| (neglecting its subscripts and superscripts) up to and including the 
fraction involving the differential expression (which may be replaced with $1$).
In cases where the differential expression is fully to the right of the argument, 
then it is a {\it termination symbol}.
Note that some scientists use an alternate notation for integrals where the
differential expression appears immediately to the right of the integral, 
e.g., $\int\!\mathrm{d}x f(x)$.
However, this notation does not appear in the DLMF. 
If such notations are encountered, we follow the same approach that we used for sums, products,
and limits (see Section~\ref{Iden4.1.2}).


\subsubsection{Extraction of variables and corresponding MEOM}
The subscripts and superscripts 
of sums, products, limits, and integrals may be different for different 
notations and are therefore challenging to parse.
For integrals, we extract the bound (integration) variable from 
the differential expression. For sums and products, the upper and lower bounds 
may appear in the subscript or superscript. Parsing subscripts is comparable 
with the problem of parsing constraints~\cite{Cohl18}
(which are often not consistently formulated).
We overcame this complexity by manually defining patterns of common 
constraints and refer to them as blueprints. 
This blueprint pattern
approach allows \LaCASt{} to identify the MEOM in the sub- and superscripts.
A more detailed explanations with examples about the blueprints is available in the 
Appendix~\ref{app:blueprints}. 

\subsubsection{Identification of operator arguments}
\label{Iden4.1.2}
Once we have extracted the bound variable for sums, products, and limits, we need to determine the end of the argument. We analyzed all sums in the DLMF and developed a heuristic that covers all the formulae in the DLMF and potentially a large portion of general mathematics. Let $x$ be the extracted bound variable. For sums, we consider a summand as a part of the argument if
   (I) it is the very first summand after the operation; or
   (II) $x$ is an element of the current summand; or
   (III) $x$ is an element of the following summand (subsequent to the current summand) and there is no termination symbol 
   between the current summand and the summand which contains $x$ with an equal or lower depth according to the parse tree (i.e., closer to the root).
We consider a summand as a single logical construct since addition and subtraction are granted a lower
operator precedence than multiplication in mathematical expressions.
Similarly, parentheses are granted higher precedence and, thus, a sequence wrapped in parentheses is part of the argument if it obeys the rules (I-III).
Summands, and such sequences, are always entirely part of sums, products, and limits or entirely not.

A termination symbol always marks the end of the argument list.
Termination symbols are relation symbols, e.g., $=$, $\neq$, $\leq$, closing parentheses or brackets, e.g., $)$, $]$, or $>$, and other operators with MEOMs, if and only if, they define the same bound variable.
If $x$ is part of a subsequent operation, then the following operator is considered as part of the argument (as in (II)).
However, a special condition for termination symbols is that it is only a termination symbol for the current chain of arguments. Consider a sum over a fraction of sums. In that case, we may reach a termination symbol within 
the fraction. However, the termination symbol would be deeper inside the parse tree as compared to 
the current list of arguments.  Hence, we used the depth to determine if a termination symbol should be 
recognized or not. Consider an unusual notation with the binomial coefficient as an example
\begin{equation}\label{eq:many-sums-ex}
    \sum_{k=0}^{n} \binom{n}{k}\ \textcolor{Red}{=} \sum_{k=0}^{n} \frac{\prod_{m=1}^{n} m}{\prod_{m=1}^{k} m \textcolor{Green}{\prod}_{m=1}^{n-k} m}.
\end{equation}
\begin{wrapfigure}{r}{0.27\textwidth}
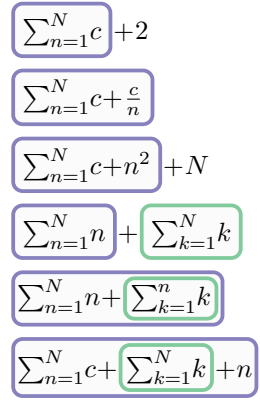

    \vspace*{-0.8cm}
    \centering
    \setlength{\belowdisplayskip}{0pt} \setlength{\belowdisplayshortskip}{0pt}
    \setlength{\abovedisplayskip}{0pt} \setlength{\abovedisplayshortskip}{0pt}
    \begin{align*}
        & \tcboxmath[boxsep=1mm]{\textstyle \sum_{n=1}^N c} + 2 \\
        & \tcboxmath[boxsep=1mm]{\textstyle \sum_{n=1}^N c + \tfrac{c}{n}}\\
        & \tcboxmath[boxsep=1mm]{\textstyle \sum_{n=1}^N c + n^2} + N \\
        & \tcboxmath[boxsep=1mm]{\textstyle \sum_{n=1}^N n} + \tcboxmath[boxsep=1mm,colframe=Green!50!white]{\textstyle \sum_{k=1}^N k} \\
        & \tcboxmath[boxsep=0.3mm]{\textstyle \sum_{n=1}^N n + \tcboxmath[boxsep=0.5mm,colback=Gray!3!white, colframe=Green!50!white, left=0mm, right=0mm,top=0mm,bottom=0mm]{\textstyle \sum_{k=1}^n k}} \\
        & \tcboxmath[boxsep=0.3mm]{\textstyle \sum_{n=1}^N c + \tcboxmath[boxsep=0.5mm,colback=Gray!3!white, colframe=Green!50!white, left=0mm, right=0mm,top=0mm,bottom=0mm]{\textstyle \sum_{k=1}^N k} + n}
    \end{align*}
    \vspace*{-0.6cm}
    \caption{Example argument identifications for sums.}
    \label{fig:ex-sum-prod}
    \vspace*{-0.8cm}
\end{wrapfigure}
This equation contains two termination symbols, marked red and green. The red termination symbol 
\textcolor{red}{$=$} is obviously for the first sum on the left-hand side of the equation. The green termination symbol \textcolor{Green}{$\prod$} terminates the product to the left because both products run over the same bound variable $m$. In addition, none of the other $=$ signs are termination symbols for the sum on the right-hand side of the equation because they are deeper in the parse tree and thus do not terminate the sum.

Note that \verb|varN| in the blueprints also matches multiple bound variable, e.g., $\sum_{m,k\in A}$.
In such cases, $x$ from above is a list of bound variables and a summand is part of the argument if one of the elements of $x$ is within this summand.
Due to the translation, the operation will be split into two preceding operations, i.e., $\sum_{m,k\in A}$ becomes $\sum_{m \in A}\sum_{k \in A}$. Figure~\ref{fig:ex-sum-prod} shows the extracted arguments for some example sums. 
The same rules apply for extraction of arguments for products and limits. 

\vspace{-0.2cm}
\subsection{Lagrange's notation for differentiation and derivatives}\label{subsec:lagrange-notation}
Another remaining issue is the Lagrange (prime) notation for 
differentiation, since it does not outwardly provide sufficient semantic information.
This notation presents two challenges.
First, we do not know with respect to which variable the differentiation 
should be performed.
Consider for example the Hurwitz zeta function $\zeta(s,a)$
\cite[\href{https://dlmf.nist.gov/25.11}{\S 25.11}]{DLMF}. In the case of 
a differentiation $\zeta'(s,a)$, it is not clear if the function should 
be differentiated with respect to $s$ or $a$.
To remedy this issue, we analyzed all formulae in the DLMF which use 
prime notations and determined which variables (slots) for which 
functions represent the variables of the differentiation.
Based on our analysis, we extended the translation patterns by meta information for semantic macros 
according to the slot of differentiation. For instance, in the case of the Hurwitz zeta function, 
the first slot is the slot for prime differentiation, i.e., 
$\zeta'(s,a) = \tfrac{\mathrm{d}}{\mathrm{d}s} \zeta(s,a)$.
The identified variables of differentiations for the special functions in the DLMF can be considered 
to be the standard slots of differentiations, e.g., in other DML, $\zeta'(s,a)$ most likely refers 
to $\tfrac{\mathrm{d}}{\mathrm{d}s} \zeta(s,a)$.

The second challenge occurs if the slot of differentiation contains complex expressions rather 
than single symbols, e.g., $\zeta'(s^2,a)$.
In this case, $\zeta'(s^2,a) = \tfrac{\mathrm{d}}{\mathrm{d}(s^2)} \zeta(s^2,a)$ instead of $\tfrac{\mathrm{d}}{\mathrm{d}s} \zeta(s^2,a)$.
Since CAS often do not support derivatives with respect to complex expressions, we use the inbuilt substitution functions\footnote{Note that \Maple{} also support an evaluation substitution via the two-argument \texttt{eval} function. Since our substitution only triggers on semantic macros, we only use \texttt{subs} if the function is defined in \Maple. In turn, as far as we know, there is no practical difference between \texttt{subs} and the two-argument \texttt{eval} in our case.} in the CAS to overcome this issue.
To do so, we use a temporary variable \verb|temp| for the substitution.
CAS perform substitutions from the inside to the outside. 
Hence, we can use the same temporary variable \verb|temp| even for nested substitutions.
Table~\ref{tab:prime-diff} shows the translation performed for $\zeta'(s^2,a)$.
CAS may provide optional arguments to calculate the derivatives for certain special functions, 
e.g., \verb|Zeta(n,z,a)| in \Map{} for the $n$-th derivative of the Hurwitz zeta function.
However, this shorthand notation is generally not supported (e.g., \Math{} does not define such 
an optional parameter). Our substitution approach is more lengthy but also more reliable.
Unfortunately, lengthy expressions generally harm the performance of CAS, especially 
for symbolic 
manipulations. Hence, we have a genuine interest in keeping translations short,
straightforward and readable. Thus, the substitution translation pattern is only 
triggered if the variable of differentiation is not a single identifier. Note that this substitution only triggers on semantic macros. Generic functions, including prime notations, are still skipped.

\begin{wraptable}{r}{0.54\textwidth}
    \centering
    \vspace{-0.8cm}
    \caption{Example translations for the prime derivative of the Hurwitz zeta function with respect to $s^2$.}
    \label{tab:prime-diff}
    \renewcommand{\arraystretch}{1.2}
    \begin{tabular}{@{}l|l@{}}
    \hline
        {\bf System} & \multicolumn{1}{c}{\bf $\zeta'(s^2,a)$} \\\hline
        {\normalsize DLMF} & { \!\verb|\Hurwitzzeta'@{s^2}{a}|}\\\hdashline
        {\normalsize \Maple{}} & { \!\verb|subs(temp=(s)^(2),diff(|}\\
        & \quad { \verb|Zeta(0,temp,a),temp$(1)))|} \\\hdashline
        {\normalsize \texttt{Mathe-}} & { \!\verb|D[HurwitzZeta[temp,a],|} \\
        {\normalsize \texttt{matica}} & \quad { \verb|{temp,1}]/.temp->(s)^(2)|}
        \\
        \hline
    \end{tabular}
    \vspace{-0.5cm}
\end{wraptable}

A related problem to MEOM of sums, products, integrals, limits, and differentiations are the 
notations of derivatives. The semantic macro for derivatives \verb|\deriv{w}{x}| (rendered as 
$\tfrac{\mathrm{d} w}{\mathrm{d} x}$) is often used with an empty first argument to render the 
function behind the derivative notation, e.g., \verb|\deriv{}{x}\sin@{x}| for 
$\tfrac{\mathrm{d}}{\mathrm{d} x}\sin x$. This leads to the same problem we faced above for 
identifying MEOMs. 
In this case,
we use the same heuristic as we 
did for sums, products, and limits. Note that derivatives may be written following 
the function argument, e.g., $\sin (x) \tfrac{\mathrm{d}}{\mathrm{d} x}$. 
If we are unable to identify any following summand that contains the variable of differentiation before we 
reach a termination symbol, we look for arguments prior to the derivative according to the heuristic (I-III).

\vspace{-0.2cm}
\subsubsection{Wronskians}
With the support of prime differentiation described above, we are also able to translate the Wronskian~\cite[\href{https://dlmf.nist.gov/1.13.E4}{(1.13.4)}]{DLMF} to \Map{} and \Math.
A translation requires one to identify the variable of differentiation from the elements of the Wronskian, e.g., $z$ for
$\mathscr{W}\{\mathrm{Ai}(z), \mathrm{Bi}(z)\}$ from~\cite[\href{https://dlmf.nist.gov/9.2.E7}{(9.2.7)}]{DLMF}.
We analyzed all Wronskians in the DLMF and discovered that most Wronskians have a special function in its argument---such as the example above.
Hence, we can use our previously inserted metadata information about the slots of differentiation to extract the variable of differentiation from the semantic macros.
If the semantic macro argument is a complex expression, we search for the identifier in the arguments that appear in both elements of the Wronskian. For example, in $\mathscr{W}\{\mathrm{Ai}(z^a), \zeta(z^2, a)\}$, we extract $z$ as the variable since it is the only identifier that appears in the arguments $z^a$ and $z^2$ of the elements.
This approach is also used when there is no semantic macro involved, i.e., from $\mathscr{W}\{z^a, z^2\}$ we extract $z$ as well.
If \LaCASt{} extracts multiple candidates or none, it throws a translation exception.

\vspace{-0.2cm}
\section{Evaluation of the DLMF using CAS}\label{sec:evaluation}
\begin{figure}[ht]
	\centering
	\vspace{-1cm}
	\includegraphics[trim=0cm 0.5cm 0cm 0cm, clip, width=\textwidth]{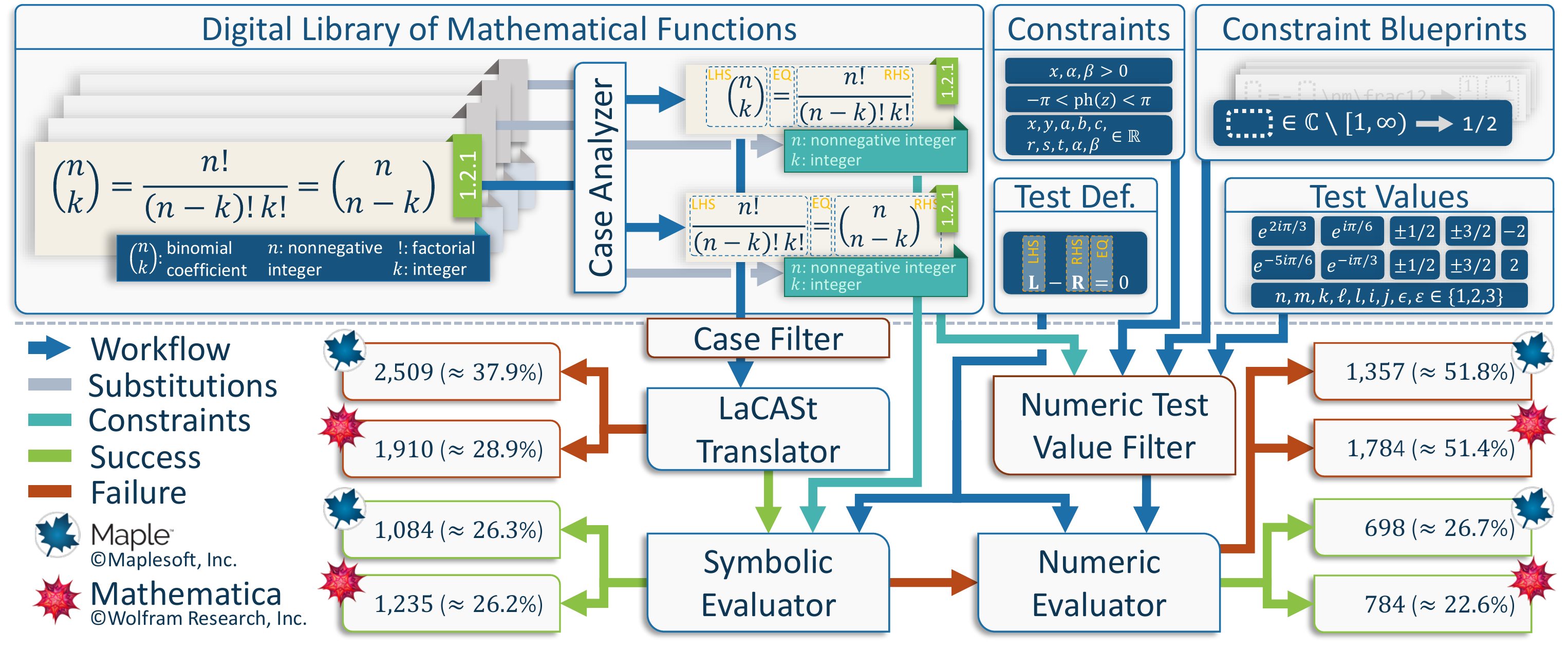}
	\vspace{-0.8cm}
	\caption{The workflow of the evaluation engine and the overall results. Errors and abortions are not included. The generated dataset contains $9,977$ equations. In total, the case analyzer splits the data into $10,930$ cases of which $4,307$ cases were filtered. This sums up to a set of $6,623$ test cases in total.}
	\label{fig:workflow}
	\vspace{-0.5cm}
\end{figure}

For evaluating the DLMF with \Maple{} and \Mathematica, we follow the same approach as demonstrated in~\cite{Cohl18}, i.e., we symbolically and numerically verify the equations in the DLMF with CAS.
If a verification fails, symbolically and numerically, we identified an issue either in the DLMF, the CAS, or the verification pipeline.
Note that an issue does not necessarily represent errors/bugs in the DLMF, CAS, or \LaCASt{} (see the discussion about branch cuts in 
Section~B).
Figure~\ref{fig:workflow} illustrates the pipeline of the evaluation engine.
First, we analyze every equation in the DLMF (hereafter referred to as test cases).
A case analyzer splits multiple relations in a single line into multiple test cases. 
Note that only the adjacent relations are considered, i.e., with $f(z) = g(z) = h(z)$, we generate two test cases $f(z) = g(z)$ and $g(z) = h(z)$ but not $f(z) = h(z)$.
In addition, expressions with $\pm$ and $\mp$ are split accordingly, e.g., $i^{\pm i} = \mathrm{e}^{\mp \pi/2}$ \cite[\href{https://dlmf.nist.gov/4.4.E12}{(4.4.12)}]{DLMF} is split into $i^{+ i} = \mathrm{e}^{- \pi/2}$ and $i^{- i} = \mathrm{e}^{+ \pi/2}$.
The analyzer utilizes the attached additional information in each line, i.e., the URL in the DLMF, the used and defined symbols, and the constraints.
If a used symbol is defined elsewhere in the DLMF, it performs substitutions.
For example, the multi-equation~\cite[\href{https://dlmf.nist.gov/9.6.E2}{(9.6.2)}]{DLMF} is split into six test cases and every $\zeta$ is replaced by $\tfrac{2}{3}z^{3/2}$ as defined in~\cite[\href{https://dlmf.nist.gov/9.6.E1}{(9.6.1)}]{DLMF}.
The substitution is performed on the parse tree of expressions~\cite{Greiner-Petter19}.
A definition is only considered as such, if the defining symbol is identical to the equation's left-hand side. That means, $z = (\tfrac{3}{2}\zeta)^{3/2}$ \cite[\href{https://dlmf.nist.gov/9.6.E10}{(9.6.10)}]{DLMF} is not considered as a definition for $\zeta$.
Further, semantic macros are never substituted by their definitions. 
Translations for semantic macros are exclusively defined by the authors.
For example, the equation~\cite[\href{https://dlmf.nist.gov/11.5.E2}{(11.5.2)}]{DLMF} contains 
the Struve $\mathbf{K}_{\nu}(z)$ function. Since \Mathematica{} does not contain this function, 
we defined an alternative translation to its definition $\mathbf{H}_{\nu}(z)-\!Y_{\nu}(z)$ in 
\cite[\href{https://dlmf.nist.gov/11.2.E5}{(11.2.5)}]{DLMF} with the Struve 
function $\mathbf{H}_{\nu}(z)$ and the Bessel function of the second kind $Y_{\nu}(z)$, because 
both of these functions are supported by \Mathematica. The second entry in 
Table~3 
in the 
Appendix~\ref{app:tables} 
shows the translation for this test case.

Next, the analyzer checks for additional constraints defined by the used symbols recursively. 
The mentioned Struve $\mathbf{K}_{\nu}(z)$ test 
case~\cite[\href{https://dlmf.nist.gov/11.5.E2}{(11.5.2)}]{DLMF} contains the Gamma function. 
Since the definition of the Gamma function~\cite[\href{https://dlmf.nist.gov/5.2.E1}{(5.2.1)}]{DLMF} 
has a constraint $\Re z >0$, the numeric evaluation must respect this constraint too. For this purpose, the case analyzer first tries to link the variables in constraints to the arguments of 
the functions. For example, the constraint $\Re z > 0$ sets a constraint for the first argument $z$ of the Gamma 
function. Next, we check all arguments in the actual test case at the same position. 
The test case contains $\Gamma(\nu + 1/2)$. In turn, the variable $z$ in the constraint of the definition of the Gamma function 
$\Re z > 0$ is replaced by the actual argument used in the test case. This adds the constraint
$\Re (\nu +1/2) > 0$ to the test case. This process is performed recursively. 
If a constraint does not contain any variable that is used in the final test case, the constraint is dropped.

In total, the case analyzer would identify four 
additional constraints for the test case~\cite[\href{https://dlmf.nist.gov/11.5.E2}{(11.5.2)}]{DLMF}. 
Table~3 
in the 
Appendix~\ref{app:tables} 
shows the applied constraints (including the directly attached constraint $\Re z > 0$ and the manually defined global constraints from Figure~\ref{fig:test-values}). Note that the 
constraints may contain variables that do not appear in the actual test case, such as $\Re \nu+k+1 > 0$. 
Such constraints do not have any effect on the evaluation because if a constraint cannot be computed to true or false, the constraint is ignored.
Unfortunately, this recursive loading of additional constraints may generate impossible conditions 
in certain cases, such as $|\Gamma(iy)|$ \cite[\href{https://dlmf.nist.gov/5.4.E3}{(5.4.3)}]{DLMF}. 
There are no valid real values of $y$ such that $\Re (iy) > 0$. In turn, every test value would be 
filtered out, and the numeric evaluation would not verify the equation. However, such 
cases are the minority 
and we were able to increase the number of correct evaluations with this feature.

To avoid a large portion of incorrect calculations, the analyzer filters the dataset before translating the test cases.
We apply two filter rules to the case analyzer. First, we filter expressions that
do not contain any semantic macros. Due to the limitations of \LaCASt, these
expressions most likely result in wrong translations. Further, it filters out several
meaningless expressions that are not verifiable, such as $z = x$ in~\cite[\href{https://dlmf.nist.gov/4.2.E4}{(4.2.4)}]{DLMF}. The result dataset flag these cases with `\textit{Skipped - no semantic math}'. Note that the result dataset still contains the translations for these cases to provide a complete picture of the DLMF. 
Second, we filter expressions that contain 
ellipsis\footnote{Note that we filter out ellipsis (e.g., \textbackslash\texttt{cdots}) but not
single dots (e.g., \textbackslash\texttt{cdot}).} (e.g., \verb|\cdots|), approximations, and 
asymptotics (e.g., $\mathcal{O}(z^2)$) since those expressions cannot be
evaluated with the proposed approach.
Further, a definition is skipped if it is not a definition of a semantic macro, such as \cite[\href{https://dlmf.nist.gov/2.3.13}{(2.3.13)}]{DLMF}, because definitions
without an appropriate counterpart in the CAS are meaningless to evaluate.
Definitions of semantic macros, on the other hand, are of special interest and remain in the test set since they allow us to test if a function in the CAS obeys the actual mathematical definition in the DLMF.
If the case analyzer (see Figure~\ref{fig:workflow}) is unable to detect a relation, i.e., split an expression on $<$, $\leq$, $\geq$, $>$, $=$, or $\neq$, the line in the dataset is also skipped because the evaluation approach relies on relations to test.
After splitting multi-equations (e.g., $\pm$, $\mp$, $a=b=c$), filtering out all non-semantic expressions, non-semantic macro definitions, ellipsis, approximations, and asymptotics, we end up with $6,623$ test cases in total from the entire DLMF. 

After generating the test case with all constraints, we translate the expression to the CAS representation. 
Every successfully translated test case is then symbolically verified, i.e., the CAS tries to simplify the difference of an equation to zero. 
Non-equation relations simplifies to Booleans. 
Non-simplified expressions are verified numerically for manually defined test values, i.e., we calculate actual numeric values for both sides of an equation and check their equivalence.


\vspace{-0.35cm}
\subsection{Symbolic Evaluation}\label{subsec:symbolic}
\vspace{-0.2cm}
The symbolic evaluation was performed for \Maple{} as in~\cite{Cohl18}. However, we use the newer 
version \Maple{} 2020. Another feature we added to \LaCASt{} is the support of packages in \Maple. 
Some functions are only available in modules (packages) that must be preloaded, such as \verb|QPochhammer| in 
the package \verb|QDifferenceEquations|\footnote{\url{https://jp.maplesoft.com/support/help/Maple/view.aspx?path=QDifferenceEquations/QPochhammer} [accessed 09/01/2021]}. 
The general \verb|simplify| method in \Maple{} does not cover $q$-hypergeometric functions. 
Hence, whenever \LaCASt{} loads functions from the $q$-hyper-geometric package, the better 
performing \verb|QSimplify| method is used.
With the WED and the new support for \Mathematica{} in \LaCASt, we perform the symbolic and numeric 
tests for \Mathematica{} as well.
The symbolic evaluation in \Mathematica{} relies on the full simplification\footnote{\url{https://reference.wolfram.com/language/ref/FullSimplify.html}\\\relax
[accessed 09/01/2021]}. 
For \Maple{} and \Mathematica, we defined the global assumptions $x,y \in \mathbb{R}$ and 
$k, n, m \in \mathbb{N}$. 
Constraints of test cases are added to their assumptions to support simplification. Adding more global assumptions for symbolic computation generally harms the performance since CAS internally uses assumptions for simplifications. It turned out that by adding more custom assumptions, the number of successfully simplified expressions decreases.

\vspace{-0.35cm}
\subsection{Numerical Evaluation}\label{subsec:numeric}
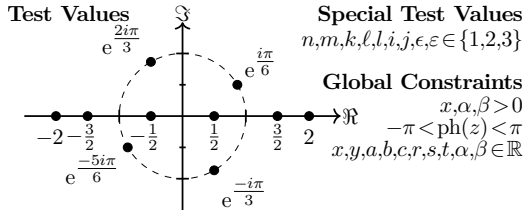
\begin{wrapfigure}{r}{0.56\textwidth}
\centering
\vspace*{-0.9cm}
\noindent\resizebox{0.565\textwidth}{!}{%
\begin{tikzpicture}
  \clip (-2.75,-1.5) rectangle (5.4, 1.76);
  \draw [thick, ->] (0,-1.5) -- (0,1.5); 
  \draw [thick, ->] (-2.5,0) -- (2.5,0); 
  \node [above] at (0,1.4) {$\Im$};
  \node [right] at (2.4,0) {$\Re$};
  
  \node [above] at (-1.8,1.4) {\textbf{Test Values}};
  
  \draw [dashed] (0, 0) circle [radius=1];
  
  \draw [thick, -] (-0.1,1) -- (0.1,1);
  \draw [thick, -] (-0.05,0.5) -- (0.05,0.5);
  \draw [thick, -] (-0.05,-0.5) -- (0.05,-0.5);
  \draw [thick, -] (-0.1,-1) -- (0.1,-1);
  
  \draw [thick, -] (-1,-0.1) -- (-1,0.1);
  \draw [thick, -] (1,-0.1) -- (1,0.1);
  
  \node [below] at (-0.6,0) {$-\tfrac{1}{2}$};
  \node [below] at (-1.6,0) {$-\tfrac{3}{2}$};
  \node [below] at (-2.1,-0.1) {$-2$};
  
  \node [below] at (0.5,0) {$\tfrac{1}{2}$};
  \node [below] at (1.5,0) {$\tfrac{3}{2}$};
  \node [below] at (2,-0.1) {$2$};
  
  \draw [fill] (-2, 0) circle [radius=2pt];
  \draw [fill] (-1.5, 0) circle [radius=2pt];
  \draw [fill] (-0.5, 0) circle [radius=2pt];
  \draw [fill] (0.5, 0) circle [radius=2pt];
  \draw [fill] (1.5, 0) circle [radius=2pt];
  \draw [fill] (2, 0) circle [radius=2pt];
  
  \node [above right] at ({cos(deg(pi/6))}, {sin(deg(pi/6))}) {$\mathrm{e}^{\tfrac{i\pi}{6}}$};
  \draw [fill] ({cos(deg(pi/6))}, {sin(deg(pi/6))}) circle [radius=2pt];
  
  \node [above left] at ({cos(deg(2*pi/3))}, {sin(deg(2*pi/3))}) {$\mathrm{e}^{\tfrac{2i\pi}{3}}$};
  \draw [fill] ({cos(deg(2*pi/3))}, {sin(deg(2*pi/3))}) circle [radius=2pt];
  
  \node [below right] at ({cos(deg(pi/3))}, {-sin(deg(pi/3))}) {$\mathrm{e}^{\tfrac{-i\pi}{3}}$};
  \draw [fill] ({cos(deg(pi/3))}, {-sin(deg(pi/3))}) circle [radius=2pt];
  
  \node [below left] at ({cos(deg(5*pi/6))}, {-sin(deg(5*pi/6))}) {$\mathrm{e}^{\tfrac{-5i\pi}{6}}$};
  \draw [fill] ({cos(deg(5*pi/6))}, {-sin(deg(5*pi/6))}) circle [radius=2pt];
  
  \node [left] at (5.5,1.6) {\textbf{Special Test Values}};
  \node [left] at (5.5,1.2) {$n, m, k, \ell, l, i, j, \epsilon, \varepsilon \in \{1,2,3\}$};
  
  \node [left] at (5.5,0.55) {\textbf{Global Constraints}};
  \node [left] at (5.5,0.15) {$x, \alpha, \beta > 0$};
  \node [left] at (5.5,-0.2) {$-\pi < \operatorname{ph}(z) < \pi$};
  \node [left] at (5.5,-0.55) {$x, y, a, b, c, r, s, t, \alpha, \beta \in \mathbb{R}$};
\end{tikzpicture}
}
\vspace{-0.9cm}
\caption{The ten numeric test values in the complex plane for general variables. The dashed line represents the unit circle $|z|=1$. 
At the right, we show the set of values for special variable values and general global constraints. On the right, $i$ is referring to a generic variable and not to the imaginary unit.}
\label{fig:test-values}
\vspace{-0.85cm}
\end{wrapfigure}

\vspace{-0.2cm}
Defining an accurate test set of values to analyze an equivalence can be an arbitrarily complex process. 
It would make sense that every expression is tested on specific values according to the containing functions.
However, this laborious process is not suitable for evaluating the entire DML and CAS.
It makes more sense to develop a general set of test values that (i) generally covers interesting domains and (ii) avoid singularities, branch cuts, and similar problematic regions.
Considering these two attributes, we come up with the ten test points illustrated in Figure~\ref{fig:test-values}. It contains four complex values on the unit circle and six points on the real axis. The test values cover the general area of interest (complex values in all four quadrants, negative and positive real values) and avoid the typical singularities at $\{0, \pm 1, \pm i\}$. 
In addition, several variables are tied to specific values for entire sections.
Hence, we applied additional global constraints to the test cases.

The numeric evaluation engine heavily relies on the performance of extracting free 
variables from an expression. Unfortunately, the inbuilt functions in CAS, if available, are not very reliable.
As the authors explained in~\cite{Cohl18}, 
a custom algorithm within \Maple{} was necessary to extract identifiers. \Mathematica{} 
has the undocumented function \verb|Reduce`FreeVariables| for this purpose. 
However, both systems, the custom solution in \Maple{} and the inbuilt \Mathematica{} 
function, have problems distinguishing free variables of entire expressions from the 
bound variables in MEOMs, e.g., integration and continuous variables.
\Mathematica{} sometimes does not extract a variable but returns the unevaluated input instead.
We regularly faced this issue for integrals. However, we discovered one example without integrals.
For \verb|EulerE[n,0]| from~\cite[\href{https://dlmf.nist.gov/24.4.E26}{(24.4.26)}]{DLMF}, 
we expected to extract $\{n\}$ as the set of free variables but instead received a set 
of the unevaluated expression itself $\{\verb|EulerE[n,0]|\}$\footnote{The bug was reported to and confirmed by Wolfram Research Version 12.0.}.
Since the extended version of \LaCASt{} handles operators, including bound variables of 
MEOMs, we drop the use of internal methods in CAS and extend \LaCASt{} to extract identifiers 
from an expression. During a translation process, \LaCASt{} tags every single identifier 
as a variable, as long as it is not an element of a MEOM. This simple approach proves 
to be very efficient since it is implemented alongside the translation process itself and 
is already more powerful as compared to the existing inbuilt CAS solutions.
We defined subscripts of identifiers as a part of the identifier, e.g., $z_1$ and $z_2$ are 
extracted as variables from $z_1 + z_2$ rather than $z$.

The general pipeline for a numeric evaluation works as follows. 
First, we replace all substitutions and extract the variables from the 
left- and right-hand sides of the test expression via \LaCASt. 
For the previously mentioned example of the Struve function~\cite[\href{https://dlmf.nist.gov/11.5.E2}{(11.5.2)}]{DLMF},
\LaCASt{} identifies two variables in the expression, $\nu$ and $z$.
According to the values in Figure~\ref{fig:test-values}, $\nu$ and $z$ are set to the general ten values.
A numeric test contains every combination of test values for all variables.
Hence, we generate $100$ test calculations for~\cite[\href{https://dlmf.nist.gov/11.5.E2}{(11.5.2)}]{DLMF}.
Afterward, we filter the test values that violate the attached constraints.
In the case of the Struve function, we end up with 25 test cases.
 
In addition, we apply a limit of 300 calculations for each test case and abort a computation after 30 seconds due to computational limitations. 
If the test case generates more than 
300 test values, only the first 300 are used.
Finally, we calculate the result for every remaining test value, i.e., we replace every variable by their value and calculate the result.
The replacement is done by \Mathematica's \verb|ReplaceAll| method because the more appropriate method \verb|With|, for unknown reasons, does not always replace all variables by their values.
We wrap test expressions in \verb|Normal| for numeric evaluations to avoid conditional expressions, which may cause incorrect calculations 
(see Section~\ref{sec:ErrorAnalysis} for a more detailed discussion of conditional outputs).
After replacing variables by their values, we trigger numeric computation.
If the absolute value of the result (i.e., the difference between left- and right-hand side of the equation) is below the defined threshold of $0.001$ or true (in the case of inequalities), the test calculation is considered successful.
A numeric test case is only considered successful if and only if every test calculation was successful.
If a numeric test case fails, we store the information on which values it failed and how many of these were successful.

\vspace{-0.4cm}
\section{Results}\label{sec:results}
\vspace{-0.2cm}
The translations to \Maple{} and \Mathematica, the symbolic results, the numeric computations, and an overview PDF of the reported bugs to \Mathematica{} are available online on our demopage.
In the following, we mainly focus on \Mathematica{} because of page limitations and because \Maple{} has been investigated more closely by~\cite{Cohl18}.
The results for \Maple{} are also available online.
Compared to the baseline ($\approx\!31\%$), our improvements doubled the amount 
translations ($\approx\!62\%$) for \Maple{} and reach $\approx\!71\%$ for \Mathematica.
The majority of expressions that cannot be translated contain macros that have no adequate translation pattern to the CAS, such as the macros for interval Weierstrass lattice 
roots~\cite[\href{https://dlmf.nist.gov/23.3.i}{\S23.3(i)}]{DLMF} 
and the multivariate hypergeometric function~\cite[\href{https://dlmf.nist.gov/19.16.9}{(19.16.9)}]{DLMF}.
Other errors ($6\%$ for \Maple{} and \Mathematica) occur for several reasons.
For example, out of the $418$ errors in translations to \Mathematica, $130$ caused an error because the MEOM of an operator could not be extracted, $86$ contained prime notations that do not refer to differentiations, $92$ failed because of the underlying \LaTeX{} parser~\cite{Youssef17},
and in $46$ cases, the arguments of a DLMF macro could not be extracted.

Out of $4,\!713$ translated expressions, $1,\!235$ ($26.2\%$) were successfully simplified by \Mathematica{} ($1,\!084$ of $4,\!114$ or $26.3\%$ in \Maple). For \Mathematica, we also count results that are equal to 0 under certain conditions as successful (called \verb|ConditionalExpression|). We identified 65 of these conditional results:
15 of the conditions are equal to constraints that were provided in the surrounding text but not in the info box of the DLMF equation; 30 were produced due to branch cut issues (see 
Section~\ref{subsec:branch-cuts}); 
and 20 were the same as attached in the DLMF but reformulated, e.g., $z \in \mathbb{C}\backslash (1, \infty)$ from \cite[\href{https://dlmf.nist.gov/25.12.E2}{(25.12.2)}]{DLMF} was reformulated to $\Im z \neq 0 \lor \Re z < 1$.
The remaining translated but not symbolically verified 
expressions were numerically evaluated for the test values in Figure~\ref{fig:test-values}.
For the $3,\!474$ cases, $784$ ($22.6\%$) were successfully verified numerically by \Mathematica{} ($698$ of $2,\!618$ or $26.7\%$ by \Maple\footnote{Due to computational issues, $120$ cases must have been skipped manually. $292$ cases resulted in an error during symbolic verification and, therefore, were skipped also for numeric evaluations. Considering these skipped cases as failures, decreases the numerically verified cases to $23\%$ in \Maple.}). For $1,\!784$ the numeric evaluation failed. 
In the evaluation process, $655$ computations timed out and $180$ failed due to errors in \Math.
Of the $1,\!784$ failed cases, $691$ failed partially, i.e., there was at least one 
successful calculation among the tested values. For $1,\!091$ all test values failed.
Table~3 
in the 
Appendix~\ref{app:tables}
shows the results for three sample test cases. The first case is a false positive evaluation because of a wrong translation. The second case is valid, but the numeric evaluation failed due to a bug in Mathematica (see next subsection). The last example is valid and was verified numerically but was too complex for symbolic verifications.

\vspace{-0.5cm}
\subsection{Error Analysis}\label{sec:ErrorAnalysis} 
\vspace{-0.1cm}
The numeric tests' performance strongly depends on the correct attached and 
utilized information. The first example in 
Table~3 
in the 
Appendix~\ref{app:tables} 
illustrates the difficulty of the task on a relatively easy case. Here, 
the argument of $f$ was not explicitly given, such as in $f(x)$. Hence, 
\LaCASt{} translated $f$ as a variable. Unfortunately, this resulted in a 
false verification symbolically and numerically. This type of error mostly 
appears in the first three chapters of the DLMF because they use 
generic functions frequently. We hoped to skip such cases by filtering 
expressions without semantic macros. Unfortunately, this derivative notation uses 
the semantic macro \verb|deriv|. In the future, we
filter expressions that contain semantic macros that are not linked to 
a special function or orthogonal polynomial.

As an attempt to investigate the reliability of the numeric test pipeline, we can run numeric evaluations on symbolically verified test cases. Since \Mathematica{} already approved a translation symbolically, the numeric test should be successful if the pipeline is reliable.
Of the $1,\!235$ symbolically successful tests, only $94$ ($7.6\%$) failed numerically. None of the failed test cases failed entirely, i.e., for every test case, at least one test value was verified. 
Manually investigating the failed cases reveal $74$ cases that failed due to an \verb|Indeterminate| response from \Mathematica{} and $5$ returned \verb|infinity|, which clearly indicates that the tested numeric values were invalid, e.g., due to testing on singularities.
Of the remaining $15$ cases, two were identical: \cite[\href{https://dlmf.nist.gov/15.9.E2}{(15.9.2)}]{DLMF} and \cite[\href{https://dlmf.nist.gov/18.5.9}{(18.5.9)}]{DLMF}. This reduces the remaining failed cases to $14$.
We evaluated invalid values for $12$ of these because the constraints for the values were given in the surrounding text but not in the info boxes.
The remaining $2$ cases revealed a bug in \Mathematica{} regarding conditional outputs (see below).
The results indicate that the numeric test pipeline is reliable, at least for relatively simple cases that were previously symbolically verified. 
The main reason for the high number of failed numerical cases in the entire DLMF ($1,\!784$) are due to missing constraints in the i-boxes and branch cut issues (see 
Section~\ref{subsec:branch-cuts}
in the Appendix), 
i.e., we evaluated expressions on invalid values.

\vspace{-0.3cm}
\subsubsection{Bug reports}
\Mathematica{} has trouble with certain integrals, which, by default, generate conditional outputs if applicable. With the method \verb|Normal|, we can suppress conditional outputs. However, it only hides the condition rather than evaluating the expression to a non-conditional output.
For example, integral expressions in~\cite[\href{https://dlmf.nist.gov/10.9.1}{(10.9.1)}]{DLMF} are automatically evaluated to the Bessel 
function $J_0(|z|)$ for the condition\footnote{$J_0(x)$ with $x \in \mathbb{R}$ is even. Hence, $J_0(|z|)$ is correct under the given condition.} $z \in \mathbb{R}$ rather than $J_0(z)$ for all $z \in \mathbb{C}$.
Setting the \verb|GenerateConditions|\footnote{\url{https://reference.wolfram.com/language/ref/GenerateConditions.html} [accessed 09/01/2021]} option to \verb|None| does not change the output. \verb|Normal| only hides $z \in \mathbb{R}$ but still returns $J_0(|z|)$.
To fix this issue, for example in \href{https://dlmf.nist.gov/10.9.1}{(10.9.1)} and \href{https://dlmf.nist.gov/10.9.4}{(10.9.4)}, we are forced to set \verb|GenerateConditions| to false.

Setting \verb|GenerateConditions| to false, on the other hand, reveals severe errors in several other cases. Consider $\int_z^{\infty} t^{-1}\mathrm{e}^{-t}\,\mathrm{d}t$ \cite[\href{https://dlmf.nist.gov/8.4.4}{(8.4.4)}]{DLMF}, which gets evaluated to $\mathrm{\Gamma}(0,z)$ but (condition) for $\Re z > 0 \land \Im z = 0$. With \verb|GenerateConditions| set to false, the integral incorrectly evaluates to $\mathrm{\Gamma}(0,z) + \ln(z)$.
This happened with the $2$ cases mentioned above.
With the same setting, the difference of the left- and right-hand sides of \cite[\href{https://dlmf.nist.gov/10.43.8}{(10.43.8)}]{DLMF} is evaluated to $0.398942$ for $x,\,\nu = 1.5$. If we evaluate the same expression on $x,\;\nu=\tfrac{3}{2}$ the result is \verb|Indeterminate| due to \verb|infinity|.
For this issue, one may use \verb|NIntegrate| rather than \verb|Integrate| to compute the integral. However, evaluating via \verb|NIntegrate| decreases the number of successful numeric evaluations in general.
We have revealed errors with conditional outputs in 
\href{https://dlmf.nist.gov/8.4.4}{(8.4.4)},
\href{https://dlmf.nist.gov/10.22.39}{(10.22.39)}, 
\href{https://dlmf.nist.gov/10.43.8}{(10.43.8-10)}, and
\href{https://dlmf.nist.gov/11.5.2}{(11.5.2)} (in \cite{DLMF}).
In addition, we identified one critical error in \Mathematica. For \cite[\href{https://dlmf.nist.gov/18.17.47}{(18.17.47)}]{DLMF}, WED (\Mathematica's kernel) ran into a \textit{segmentation fault (core dumped)} for $n > 1$. The kernel of the full version of \Mathematica{} gracefully died without returning an output\footnote{All errors were reported to and partially confirmed by Wolfram Research. See 
Appendix~\ref{app:issues} 
for more information.}.

Besides \Mathematica, we also identified several issues in the DLMF.
None of the newly identified issues were critical, such as the reported sign error from the previous project~\cite{Cohl18}, but generally refer to missing or wrong attached semantic information.
With the generated results, we can effectively fix these errors and further semantically enhance the DLMF. 
For example, some definitions are not marked as such, e.g., $Q(z) = \int_0^{\infty} \mathrm{e}^{-zt} q(t)\,\mathrm{d}t$ \cite[\href{https://dlmf.nist.gov/2.4.E2}{(2.4.2)}]{DLMF}.
In \cite[\href{https://dlmf.nist.gov/10.24.E4}{(10.24.4)}]{DLMF}, $\nu$ must be a real value but was linked to a \textit{complex parameter} and $x$ should be positive real. An entire group of cases \cite[\href{https://dlmf.nist.gov/10.19.E10}{(10.19.10-11)}]{DLMF} also discovered the incorrect use of semantic macros. In these formulae, $P_k(a)$ and $Q_k(a)$ are defined but had been incorrectly marked up as Legendre functions going all the way back to DLMF Version 1.0.0
(May 7, 2010).
In some cases, equations are mistakenly marked as definitions, e.g., \cite[\href{https://dlmf.nist.gov/9.10.E10}{(9.10.10)}]{DLMF} and \cite[\href{https://dlmf.nist.gov/9.13.E1}{(9.13.1)}]{DLMF} are annotated as local definitions of $n$.
We also identified an error in \LaCASt, which incorrectly translated the exponential integrals $E_1(z)$, $\operatorname{Ei}(x)$ and $\operatorname{Ein}(z)$ (defined in \cite[\href{https://dlmf.nist.gov/6.2.i}{\S6.2(i)}]{DLMF}). 
A more explanatory overview of discovered, reported, and fixed issues in the DLMF, \Mathematica, and \Maple{} is provided in the 
Appendix~\ref{app:issues}.

\vspace{-0.4cm}
\section{Conclusion}\label{sec:future-work}
\vspace{-0.3cm}
We have presented a novel approach to verify the theoretical digital mathematical library DLMF with the power of two major general-purpose computer algebra systems \Maple{} and \Mathematica. With \LaCASt, we transformed the semantically enhanced \LaTeX{} expressions from the DLMF to each CAS. Afterward, we symbolically and numerically evaluated the DLMF expressions in each CAS. Our results are auspicious and provide useful information to maintain and extend the DLMF efficiently. We further identified several errors in \Mathematica, \Maple{}~\cite{Cohl18}, the DLMF, and the transformation tool \LaCASt, proving the profit of the presented verification approach.
Further, we provide open access to all results, including translations and evaluations\footnote{\demolink\ [accessed 01/01/2022]}. 
and to the source code of \LaCASt\footnote{\url{https://github.com/ag-gipp/LaCASt} [accessed 04/01/2022]}.

The presented results show a promising step towards an answer for our initial research question. By translating an equation from a DML to a CAS, automatic verifications of that equation in the CAS allows us to detect issues in either the DML source or the CAS implementation. Each analyzed failed verification successively improves the DML or the CAS. Further, analyzing a large number of equations from the DML may be used to finally verify a CAS. In addition, the approach can be extended to cover other DML and CAS by exploiting different translation approaches, e.g., via \MathML~\cite{SchubotzGSMCG18} or OpenMath~\cite{HerasPR09}.

Nonetheless, the analysis of the results, especially for an entire DML, is cumbersome. Minor missing semantic information, e.g., a missing constraint or not respected branch cut positions, leads to a relatively large number of false positives, i.e., unverified expressions correct in the DML and the CAS. This makes a generalization of the approach challenging because all semantics of an equation must be taken into account for a trustworthy evaluation.
Furthermore, evaluating equations on a small number of discrete values will never provide sufficient confidence to verify a formula, which leads to an unpredictable number of true negatives, i.e., erroneous equations that pass all tests.
A more sophisticated selection of critical values or other numeric tools with automatic results verification (such as variants of Newton's interval method) potentially mitigates this issue in the future.
After all, we conclude that the approach provides valuable information to complement, improve, and maintain the DLMF, \Maple, and \Mathematica.
A trustworthy verification, on the other hand, might be out of reach.

\vspace{-0.3cm}
\subsection{Future Work}
\vspace{-0.1cm}
The resulting dataset provides valuable information about the differences between CAS and the DLMF. These differences had not been largely studied in the past and are worthy of analysis. Especially a comprehensive and machine-readable list of branch cut positioning in different systems is a desired goal~\cite{Corless00}. Hence, we will continue to work closely together with the editors of the DLMF to improve further and expand the available information on the DLMF. Finally, the numeric evaluation approach would benefit from test values dependent on the actual functions involved. For example, the current layout of the test values was designed to avoid problematic regions, such as branch cuts. However, for identifying differences in the DLMF and CAS, especially for analyzing the positioning of branch cuts, an automatic evaluation of these particular values would be very beneficial and can be used to collect a comprehensive, inter-system library of branch cuts. Therefore, we will further study the possibility of linking semantic macros with numeric regions of interest.

{
%
\section*{Acknowledgements} 
We thank J\"urgen Gerhard from Maplesoft for providing access and support for Maple. We also thank the DLMF editors for their assistance and support.
This work was supported by the German Research Foundation (DFG grant no.: GI 1259/1) and the German Academic Exchange Service (DAAD grant no.: 57515245).
}

\printbibliography


\pagebreak

{\small\medskip\noindent{\bf Open Access} This chapter is licensed under the terms of the Creative Commons\break Attribution 4.0 International License (\url{http://creativecommons.org/licenses/by/4.0/}), which permits use, sharing, adaptation, distribution and reproduction in any medium or format, as long as you give appropriate credit to the original author(s) and the source, provide a link to the Creative Commons license and indicate if changes were made.}

{\small \spaceskip .28em plus .1em minus .1em The images or other
third party material in this chapter are included in the\break
chapter's Creative Commons license, unless indicated otherwise in a
credit line to the\break material.~If material is not included in
the chapter's Creative Commons license and\break your intended use
is not permitted by statutory regulation or exceeds the
permitted\break use, you will need to obtain permission directly
from the copyright holder.}

\medskip\noindent\includegraphics{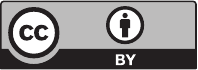}

\newpage

\appendix

\begin{center}
\Huge Appendix
\end{center}

\section{MEOM Blueprints}\label{app:blueprints}

\begin{wraptable}[18]{r}{6.3cm}
    \centering
    \vspace{-0.7cm}
    \caption{The table contains examples of the blueprints for subscripts of sums/products including an example expression that matches the blueprint.}
    \label{tab:blueprint-sum-prod}
    \renewcommand{\arraystretch}{1.2}
    \begin{tabular}{ r | c }
        \hline
        \multicolumn{1}{c|}{\textbf{Blueprints}} & \textbf{Example} \\\hline
        \makecell[r]{\texttt{numL1 \textbackslash leq var1 <}\\\texttt{var2 \textbackslash leq numU1}}
        & $0 \!\leq\! n \!<\! k \!\leq\! 10$ \\\hdashline
        \verb|-\infty < varN < \infty|      & $-\infty \!<\! n \!<\! \infty$ \\\hdashline
        \verb|numL1 < varN < numU1|         & $0 \!<\! n, k \!<\! 10$ \\\hdashline
        \verb|numL1 \leq varN < numU1|      & $0 \!\leq\! k \!<\! 10$ \\\hdashline
        \verb|numL1 < varN \leq numU1|      & $0 \!<\! n, k \!\leq\! 10$ \\\hdashline
        \verb|varN \leq numU1|              & $n, k \!\leq\! N+5$ \\\hdashline
        \verb|varN \in numL1|               & $n \!\in\! \{1,2,3\}$ \\\hdashline
        \verb|varN = numL1|                 & $n,k,l \!=\! 1$ \\\hline
    \end{tabular}
    \vspace{-0.28cm}
\end{wraptable}

In this section, we briefly explain the MEOM blueprints. Those blueprints are mathematical expressions with wild cards which are tied to a specific rule-based interpretations.
If one of our blueprints matches an expression, we identified the necessary MEOM elements, i.e., the argument(s) and bound variable(s) of the mathematical operators.

For our MEOM blueprints, we define three placeholders (wild cards): \verb|varN| for single 
identifiers or a list of identifiers (delimited by commas), \verb|numL1|, 
and \verb|numU1|, representing lower and upper bound expressions, 
respectively. 
In addition, for sums and products, we need to distinguish between including and excluding boundaries, e.g., $1 \!<\! k$ and $1 \!\leq\! k$.
An excluding relation, such as $0\!<\!k\!<\!10$, must be interpreted as a sum from $1$ to $9$.
Table~\ref{tab:blueprint-sum-prod} shows the final set of sum/product subscript blueprints.

Standard notations may not explicitly show infinity boundaries. Hence, we set the default boundaries to infinity.
For limit expressions we need different blueprints to capture the limit direction. 
We cover the standard notations with
`\verb|var1 \to numL*|',
where \verb|*| is either \verb|+|, \verb|-|, \verb|^+|, \verb|^-| or absent and the different arrow-notations where \verb|\to| can be either \verb|\downarrow|, \verb|\uparrow|, \verb|\searrow|, or \verb|\nearrow|, specifying one-sided limits.
Note that the arrow-notation (besides \verb|\to|) is not used in the DLMF and thus, has no effect on the performance of \LaCASt{} in our evaluation.

The blueprint approach can be easily extended to new patterns, 
which helps to maintain \LaCASt{} and support more expressions.
In fact, the blueprint approach is flexible enough to parse more complex situations, 
such as multi-line subscript expressions. 
However, there are scenarios in which the blueprint approach is not enough to perform a translation.
Consider the divisor sum $\sum_{(p-1) | 2n} 1/p$ \cite[\href{https://dlmf.nist.gov/24.10.E1}{(24.10.1)}]{DLMF}, 
where the sum is over all $p$ such that $p-1$ divides $2n$. A proper translation needs 
to acknowledge that $p-1$ rather than $p$ divides $2n$. Hence, a translation to \Mathematica{} 
potentially manipulates the $p$ in the argument of the sum, to adjust this. A proper 
translation could be \verb|Sum[1/(p+1), {p, Divisors[2*n]}]|. 
However, such manipulations quickly increase in complexity and require symbolic 
computation when reaching a certain level.
This is currently out of scope for \LaCASt.
Note that blueprints could also cover several scenarios with ellipsis, such as in
$\sum_{1 < n_1 < \cdots < n_k < m}$.
However, a proper analysis of expressions with ellipsis is still an open issue for \LaCASt.

\section{Why Branch Cuts Matter}\label{subsec:branch-cuts}

Problems that we regularly faced during evaluation are issues related to multi-valued functions. 
Multi-valued functions map values from a domain to multiple values in a codomain and frequently appear in the complex analysis of elementary and special functions. Prominent examples are the inverse trigonometric functions, the complex logarithm, or the square root. A proper mathematical description of multi-valued functions requires the complex analysis of Riemann surfaces. Riemann surfaces are one-dimensional complex manifolds associated with a multi-valued function. One usually multiplies the complex domain into a many-layered covering space. 
The correct properties of multi-valued functions on the complex plane may no longer be valid by their counterpart functions on CAS, e.g., $(1/z)^w$ and $1/(z^w)$ for $z,w \in \mathbb{C}$ and $z \neq 0$.
For example, consider $z,w \in \mathbb{C}$ such that $z \neq 0$. Then mathematically, $(1/z)^w$ always equals $1/(z^w)$ (when defined) for all points on the Riemann surface with fixed $w$. However, this should certainly not be assumed to be true in CAS, unless very specific assumptions are adopted (e.g., $w \in \mathbb{Z}, z > 0$). For all modern CAS\footnote{The authors are not aware of any example of a CAS which treats multi-valued functions without adopting principal branches.}, this equation is not true. Try, for instance, $w = 1/2$. Then $(1/z)^{1/2} - 1/z^{1/2} \neq 0$ on CAS, nor for $w$ being any other rational non-integer number.

The resulting ranges of multi-valued functions are referred to as branches, and the curves which separate these branches are called branch cuts. The restricted range which is associated with
the range typically adopted using real numbers, is often referred to
as the principal branch. 
In order to compute multi-valued functions, CAS choose branch cuts for these functions so that they may evaluate them on their principal branches.
Branch cuts may be positioned differently among CAS~\cite{Corless00}, e.g., $\operatorname{arccot}(-\frac12) \approx 2.03$ in \Maple{} but is $\approx -1.11$ in \Mathematica. This is certainly not an error and is usually well documented for specific CAS~\cite{MultivaluedCAS,MapleBranch}.
However, there is no central database that summarizes branch cuts in different CAS or DML. The DLMF as well, explains and defines their branch cuts carefully but does not carry the information within the info boxes of expressions. 
Due to complexity, it is rather easy to lose track of branch cut positioning and evaluate expressions on incorrect values. For example, consider the equation~\cite[\href{https://dlmf.nist.gov/12.7.10}{(12.7.10)}]{DLMF}. A path of $z(\phi) = e^{i\phi}$ with $\phi \in [0, 2\pi]$ would pass three different branch cuts. An accurate evaluation of the values of $z(\phi)$ in CAS require calculations on the three branches using analytic continuation. 
\LaCASt{} and our evaluation frequently fall into the same trap by evaluating values that are no longer on the principal branch used by CAS.
To solve this issue, we need to utilize branch cuts not only for every function but also for every equation in the DLMF~\cite{Greiner-Petter19}.
The positions of branch cuts are exclusively provided in the text but not in the i-boxes.
Adding the information to each equation in the DLMF would be a laborious process because a branch cut position may change according to the used values (see the example~\cite[\href{https://dlmf.nist.gov/12.7.10}{(12.7.10)}]{DLMF} from above).
Our result data, however, would provide beneficial information to update, extend, and maintain the DLMF, e.g., by adding the positions of the branch cuts for every function.

\section{Overview of Bug Reports and Discovered Issues}\label{app:issues}

Throughout the development of \LaCASt{} and especially during the research on this paper, we identified several issues in the DLMF, \Maple, and \Mathematica.
Some of these issues are severe while most of them are minor problems.
With this section, we want to take the opportunity to conclude the progress of \LaCASt{} as a verification approach and summarize the more prominent issues we discovered over the time.
Please note that some of these issues (especially in regard of the DLMF and \Maple) have been reported before and even published in previous publications.

\subsection{Digital Library of Mathematical Functions}
Since \LaCASt{} was always developed in collaboration with developers of the DLMF, numerous of minor fixes, tweaks, and updates have been implemented over the time.
Most of them are not worth noting with a few exceptions.
The first error in the DLMF that we discovered with the help of \LaCASt~\cite{Greiner-Petter19} was the sign error in~\dlmf{14.5.14}
\begin{equation}
\mathrm{Q}_{\nu}^{-1/2}\!(\cos\ \theta) = \textcolor{Red}{\mathbf{-}} \left(\frac{\pi}{2\sin\ \theta}\right)^{1/2}\ \frac{\cos\left(\left(\nu+\tfrac{1}{2}\right)\,\theta\right)}{\nu+\tfrac{1}{2}}.
\end{equation}
This error also appeared in the original \textit{Handbook of Special Functions}~\cite[p.~359]{Olver2010} and was fixed with DLMF version 1.0.16 in September 2017.

An entire group of equations \cite[\href{https://dlmf.nist.gov/10.19.E10}{(10.19.10-11)}]{DLMF} used semantic macros incorrectly and therefore yielded to wrong links and annotations visible in the attached information box next to the equation in the DLMF. In these formulae, $P_k(a)$ and $Q_k(a)$ are defined but had been incorrectly marked up as Legendre functions going all the way back to DLMF version 1.0.0.
This error has been fixed due to our feedback with DLMF version 1.0.27 in June 2020.

Minor discovered issues include a missing comma in the constraint~\dlmf{10.16.7} $2\nu\neq -1, -2\textcolor{Red}{\mathbf{,}} -3, \ldots$ which was also missing in the DLMF book~\cite[p.~228]{Olver2010} (fixed with v. 1.0.19), unmarked~\dlmf{2.4.2} or erroneously marked definitions in~\dlmf{9.10.10} and in~\dlmf{9.13.1} (all remain unsolved), and wrong annotations of $\nu$ as \textit{complex parameter} and $x$ as \textit{real} while \textit{real value} and \textit{positive real}, respectively, would be correct in~\dlmf{10.24.4} (remain unsolved).
Additionally, due to \LaCASt{}, the ambiguous semantic macro \verb|\Wron| for Wronskians has been revised so that the variable which is differentiated against is precisely specified in 72 occasions~\cite{Cohl18}.

\subsection{Maple}
Via \LaCASt{}, we discovered a bug in \Maple's 2016 \verb|simplify| procedure.
For the equation~\dlmf{7.18.4}
\begin{equation}\label{eq:erfc}
\frac{\mathrm{d}^n}{\mathrm{d}z^n} \left(e^{z^{2}}\operatorname{erfc}\, z\right)=(-1)^{n}2^{n}n!e^{z^{2}} \mathrm{i}^n \operatorname{erfc}\!(z), \quad n = 0, 1, 2, \ldots,
\end{equation}
where $e$ is the base of the natural logarithm, $\operatorname{erfc}(z)$ is the complementary error function, and $\mathrm{i}^n\operatorname{erfc}(z)$ the repeated integrals of the complementary error function, \LaCASt{} correctly generated the following translation:
\begin{translationbox}{\LaCASt{} translation of equation~(\ref{eq:erfc}) to \Maple}
\begin{mycode}[language=myMa, mathescape=false]
diff( exp(z^2)*erfc(z), [z$(n)] ) = (-1)^(n)*(2)^(n)*factorial(n)*exp(z^2)*erfc(n, z)
\end{mycode} 
\vspace{-0.2cm}
\begin{footnotesize}
\hfill\rule{0.4\textwidth}{.4pt}

\vspace{-0.13cm}
\hfill\footnotesize Redundant parentheses removed to improve readability.
\end{footnotesize}
\end{translationbox}

\begin{figure}[ht]
	\centering
	\vspace*{-0.5cm}
	\includegraphics[width=0.55\textwidth]{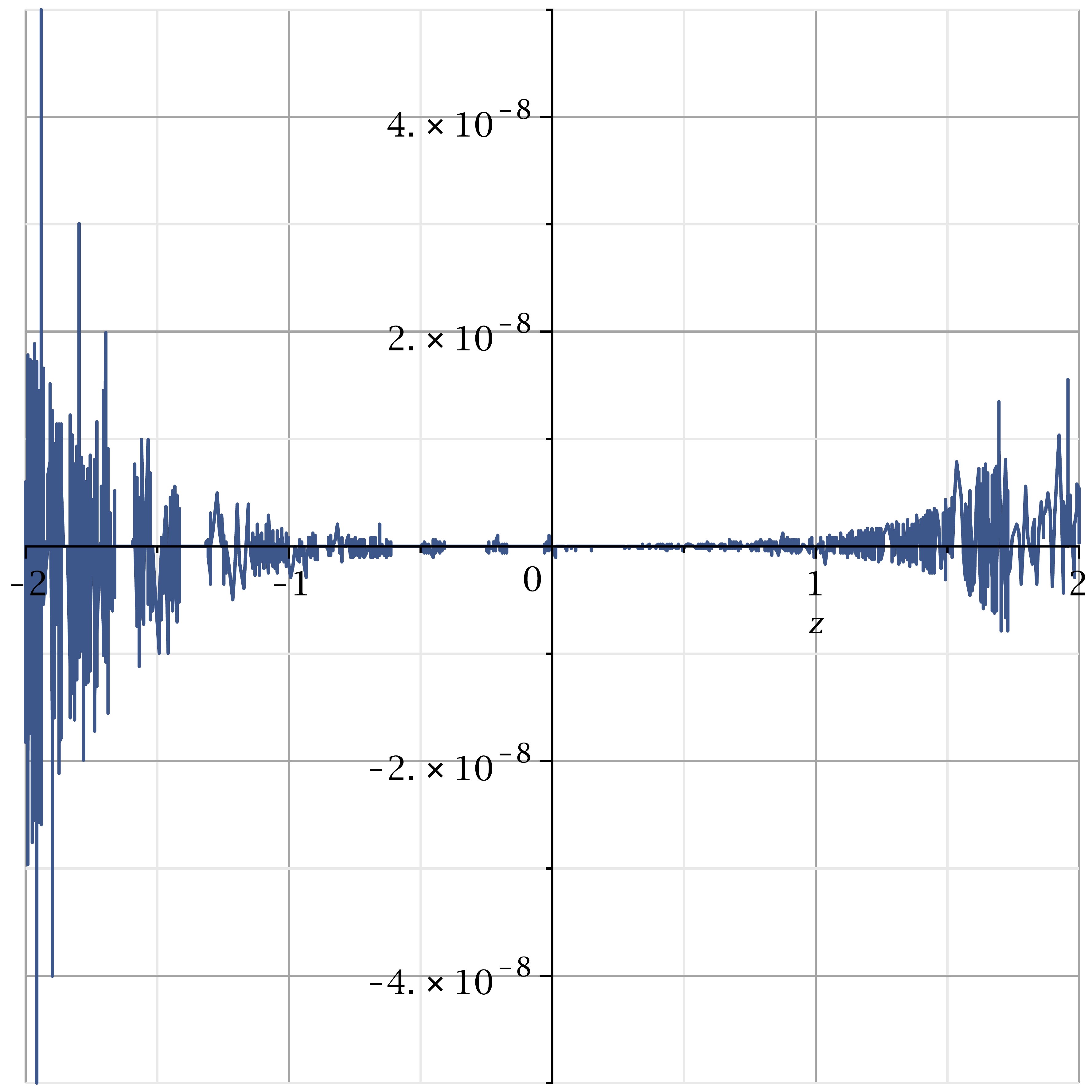}
	\vspace*{-0.5cm}
	\caption{The difference of the left- and right-hand side of equation~(\ref{eq:erfc}) evaluated in \Maple{} for $n=0$ and $z \in [-2, 2]$.}
	\label{fig:maple-plot}
	\vspace*{-0.5cm}
\end{figure}

\Maple{} 2016 falsely returns $0$ when we call the \verb|simplify| procedure for the translated left-hand side of the equation.
Maplesoft has confirmed this defect in \Maple{} 2016 in private communications~\cite{Greiner-Petter19}.
Although an updated behavior occurred in \Maple{} 2018 and 2020, the error still persists today.
\Maple{} version 2020.2 automatically evaluates the left-hand side of equation~(\ref{eq:erfc}) to the rather complex expression
\begin{equation}
\begin{split}
& \sum_{k=0}^n \frac{1}{\sqrt{\pi}} \binom{n}{k} \left( \sum_{m=0}^k e^{z^2} B_{k,m}  \left( (2)_1 z, \ldots, (2-k+m)_{k-m+1} z^{1-k+m} \right) \right) \\
& \left( -G_{2,3}^{1,2} \left( z; \genfrac{}{}{0pt}{}{0, \tfrac{1}{2}}{0, -\tfrac{1}{2}+\tfrac{n}{2}-\tfrac{k}{2}, \tfrac{n}{2}-\tfrac{k}{2}} \right) 2^{n-k} + (1-n+k)_{n-k} \sqrt{\pi} z^{n-k-1} \right) z^{1-n+k},
\end{split}
\end{equation}
where $G_{p,q}^{m,n}\left( z; \genfrac{}{}{0pt}{}{a_1, \ldots, a_p}{b_1, \ldots, b_q} \right)$ is the Meijer $G$-function~\dlmf{16.17.1}, $B_{n,k} (x_1, \ldots, x_{n-k+1})$ the incomplete Bell polynomials~\cite{Bell34}, and $(x)_{n}$ the Pochhammer's symbol~\dlmf{5.2.4}.
For small $n$ and $z$, the difference of left- and right-hand side of equation~(\ref{eq:erfc}) is indeed almost zero up to the machine accuracy.
For large absolute values of $z$, however, the difference increases quickly.
Figure~\ref{fig:maple-plot} plots the difference of left- and right-hand side of equation~(\ref{eq:erfc}) for $n=0$ and $z \in [-2,2]$.
For $|z| > 4$, the difference is already larger then $1.0$. 
This might be produced by accumulated round-off errors because smaller values are calculated with greater precision in floating-point arithmetics.

\subsection{Mathematica}

As we pointed out in Section~\ref{sec:ErrorAnalysis}, we discovered some trouble with integrals in \Mathematica{} and confusing behavior with rational numbers.
After discussing these cases with \Mathematica{} developers, some of them have been confirmed as bugs. 
Other cases, however, were the results of our testing methodology. 
First, we take a look at the confirmed errors.
The most crucial report was about~\dlmf{18.17.14}
\begin{equation}\label{eq:bug-equation}
\frac{x^{\alpha+\mu}L_n^{(\alpha+\mu)}(x)}{\mathrm{\Gamma}(\alpha + \mu + n + 1)}=\int_{0}^{x}\frac{y^{\alpha}L_n^{(\alpha)}(y)}{\mathrm{\Gamma}(\alpha + n + 1)}\frac{(x-y)^{\mu-1}}{\mathrm{\Gamma}(\mu)}\mathrm{d}y.
\end{equation}
For this equation, we calculated the difference of the left- and right-hand side as usual
\begin{equation}
\frac{x^{\alpha+\mu}L_n^{(\alpha+\mu)}(x)}{\mathrm{\Gamma}(\alpha + \mu + n + 1)} \textcolor{Red}{-} \int_{0}^{x}\frac{y^{\alpha}L_n^{(\alpha)}(y)}{\mathrm{\Gamma}(\alpha + n + 1)}\frac{(x-y)^{\mu-1}}{\mathrm{\Gamma}(\mu)}\mathrm{d}y
\end{equation}
and computed numerical test values for this difference.
In particular, we identified the four variables $x,\ n,\  \alpha$, and $\mu$.
As described in Figure~\ref{fig:test-values} (Section~\ref{subsec:numeric}) in the paper, $n$ is defined as a special variable bind to the numeric values $\{1,2,3\}$, $x$ and $\alpha$ are positive real values of our general test values, i.e., $x, \alpha \in \{ \tfrac{1}{2}, \tfrac{3}{2}, 2 \}$, and $\mu$ is not further limited, i.e., $\mu \in \left\{ \pm \tfrac{1}{2}, \pm \tfrac{3}{2}, \pm 2, e^{\tfrac{i\pi}{6}}, e^{\tfrac{2i\pi}{3}}, e^{\tfrac{-i\pi}{3}}, e^{\tfrac{-5i\pi}{6}} \right\}$.
This resulted in $270$ test value combinations which are further limited by the attached (local) constraints in the DLMF~\dlmf{18.17.14}: $\mu > 0, x > 0$.
Since $x$ was already constraint to positive real values with our global constraints, the second local constraint has no additional effect. For $\mu$, we must note that we also removed invalid comparisons. For example, $e^{\tfrac{2i\pi}{3}} > 0$ throws an error in \Mathematica{} due to the invalid comparison with an imaginary number. Hence, $\mu > 0$ filtered out imaginary numbers and negative real values. Here, we ended up with $\{ \tfrac{1}{2}, \tfrac{3}{2}, 2 \}$.
Additionally, we see equation~\ref{eq:bug-equation} contains $\mathrm{\Gamma}(\alpha+\mu+n+1)$, $\mathrm{\Gamma}(\alpha+n+1)$, and $\mathrm{\Gamma}(\mu)$. The definition of the Gamma function in the DLMF~\dlmf{5.2.1} contains the additional constraint $\Re z > 0$ where $z$ is the argument of the Gamma function. Hence, we retrieved three additional constraints for equation~\ref{eq:bug-equation}: $\Re \left( \alpha+\mu+n+1 \right) > 0$, $\Re \left( \alpha+n+1 \right) > 0$, and $\Re \left( \mu \right) > 0$. However, none of the constraints further reduced our test value set.
Conclusively, we end up with $81$ ($= 3^4$) test value combinations: $n \in \{1,2,3\}$ and $x, \alpha, \mu \in \{ \tfrac{1}{2}, \tfrac{3}{2}, 2 \}$.

Interestingly, for rational inputs (fractions) of test values, the difference was zero.
For example, evaluating the difference on $x,\;a,\;\mu = \tfrac{3}{2}$ and $n=2$ returns zero.
However, the same test with floating point numbers, i.e., $x,\;a,\;\mu\;=\;1.5$ and $n\;=\;2$, result in a fatal segmentation fault\footnote{A \textit{segmentation fault} is an access violation of protected memory. For example, an operating system can prevent a program \texttt{A} to change the memory of another program \texttt{B} and sends a \textit{segmentation fault} to \texttt{A}. This signal generally causes program \texttt{A} to abnormally terminate, unless special error handling was implemented.} causing \Mathematica{} to crash.
\begin{translationbox}{Segmentation Fault Example in \Mathematica{} v. 12.1.1}
\begin{mycode}[language=myMMA, mathescape=false]
In[1] := expr = -Integrate[((x - y)^(-1 + mu)*y^a*LaguerreL[n, a, y])/ (Gamma[1 + n + a]*Gamma[mu]), {y, 0, x}] + (x^(a + mu)*LaguerreL[n, a + mu, x])/ Gamma[1 + n + a + mu];

In[2] := ReplaceAll[expr, {n -> 2, x -> 3/2, a -> 3/2, mu -> 3/2}] 
Out[2] = 0

In[3] := ReplaceAll[expr, {n -> 2, x -> 1.5, a -> 1.5, mu -> 1.5}] 
Segmentation fault (core dumped)
\end{mycode}
\end{translationbox}

This issue was reported\footnote{Case ID: 4664157} and later fixed\footnote{The fix was communicated to us via a new case ID: 4776927} with \Mathematica{} version 12.2 (released November 2020).
The developers told us this issue can be traced back to version 10.4 (released March 2016).

We further identified errors in the variable extraction procedure in \Mathematica.
For example, for~\dlmf{24.4.26}
\begin{equation}
E_n(0) = -E_n(1)=-\frac{2}{n+1}(2^{n+1}-1)B_{n+1},
\end{equation}
we expected to extract just $n$ as the free variable.
We reduced the issue to a minimal working example just for the most left-hand side of the equation.
\begin{translationbox}{False Variable Extraction in \Mathematica{} v. 12.0}
\begin{mycode}[language=myMMA, mathescape=false]
In[1] := Reduce`FreeVariables[EulerE[n, 0]]
Out[1] = {EulerE[n, 0]}
\end{mycode}
\end{translationbox}
This particular error was confirmed and has been fixed\footnote{Case ID: 4373302}.
However, since the procedure \verb|Reduce`FreeVariables| is not a publicly documented function in \Mathematica, the method remain unstable.
Especially in mathematical operators with bounded variables, such as sums, products, integrals, and limits, the procedure tend to generate inaccurate results.

In regard of the outlined issues with the \verb|GenerateConditions| flag in integrals, most problematic cases were the result of using \verb|ReplaceAll| to set numeric values for variables.
Consider, for example~\dlmf{10.43.8}
\begin{equation}
\int_{0}^{x}e^{\pm t}t^{-\nu}I_{\nu}(t)\mathrm{d}t=-\frac{e^{\pm x}x^{-\nu+1}}{2\nu-1}(I_{\nu}(x)\mp I_{\nu-1}(x))\mp \frac{2^{-\nu+1}}{(2\nu-1)\mathrm{\Gamma}(\nu)}.
\end{equation}
First, \LaCASt{} splits the expression in two test cases by resolving $\pm$ and $\mp$.
For the first case, i.e., $\pm$ is replaced by $+$ and $\mp$ by $-$, \Mathematica{} automatically evaluates the test expression, i.e., the difference of left- and right-hand side of the equation: 
\begin{equation}\label{eq:modbessel}
\left(\int_{0}^{x}e^{t}t^{-\nu}I_{\nu}(t)\mathrm{d}t\right) \textcolor{Red}{-} \left( -\frac{e^{x}x^{-\nu+1}}{2\nu-1}(I_{\nu}(x)-I_{\nu-1}(x))-\frac{2^{-\nu+1}}{(2\nu-1)\mathrm{\Gamma}(\nu)} \right)
\end{equation}
with the \verb|GenerateConditions| set to \verb|None| for the integral to
\begin{align}
\begin{split}
& \frac{e^x x^{-\nu} \left( -x + 2\nu\right) I_{\nu}(x)}{-1+2\nu} + 
\frac{e^x x^{1-\nu}\left( -I_{-1+\nu}(x) + I_{\nu}(x) \right)}{-1+2\nu} + \\
& \frac{e^x x^{1-\nu} I_{1+\nu}(x)}{-1+2\nu} +
\frac{2^{1-\nu}}{(-1+2\nu) \mathrm{\Gamma}(z)} +
\frac{2^\nu \sqrt{\pi} \nu \operatorname{sec}(\pi \nu)}{\mathrm{\Gamma}\left(\tfrac{3}{2}-\nu\right) \mathrm{\Gamma}(1+2\nu)}.
\end{split}
\end{align}
This happens because CAS automatically perform some computations on their inputs unless we prevent it (e.g., via \verb|Hold|). However, evaluating this expression now on $x,\;\nu = 1.5$ returns $0.398942$ rather than the expected zero.
For $x,\;\nu = \tfrac{3}{2}$, the return value is infinity (i.e., indeterminate).
The issue was acknowledged by the developers, who explained that the last term causes the behaviour, because 
\begin{equation}
\frac{2^\nu \sqrt{\pi}\; \nu\; \operatorname{sec}(\pi \nu)}{\mathrm{\Gamma}\left(\tfrac{3}{2}-\nu\right) \mathrm{\Gamma}(1+2\nu)}
\end{equation}
is \verb|Infinity/Infinity| for $\nu = 3/2$.
A workaround to this issue is to use \verb|Limit| rather than \verb|ReplaceAll| to evaluate the expression on specific values.
\begin{translationbox}{Limit Workaround for equation~\ref{eq:modbessel}}
\begin{mycode}[language=myMMA, mathescape=false]
In[2] := N[ReplaceAll[expr,{Rule[x,3/2], Rule[nu,3/2]}]]
Out[2] = Indeterminate

In[3] := Limit[ expr, {Rule[x,3/2], Rule[nu,3/2]} ]
Out[3] = 0
\end{mycode}
\vspace{-0.2cm}
\begin{footnotesize}
\hfill\rule{0.4\textwidth}{.4pt}

\vspace{-0.13cm}
\hfill\footnotesize \texttt{expr} is the input of equation~(\ref{eq:modbessel}).
\end{footnotesize}
\end{translationbox}
To the best of our knowledge, this \textit{issue} still persists\footnote{As of 9/9/2021}. 
If this behavior is intended (or even desired) is up for debate.
Yet, it is another characteristic of CAS to keep track of.
The same workaround was suggested for~\dlmf{11.5.2} and~\dlmf{11.5.8-10}.
In case of~\dlmf{10.9.1}
\begin{equation}
J_0(z)=\frac{1}{\pi}\int_{0}^{\pi}\cos (z\;\cos\; \theta)\mathrm{d}\theta,
\end{equation}
the right-hand side of the equation was evaluated to \verb|BesselJ[0,z]| by \Mathematica{} for \verb|GenerateConditions -> False|. Which is correct. However, without this flag (or set to \verb|None|), \Mathematica{} returns \verb|BesselJ[0, Abs[z]]| if $z \in \mathbb{R}$.
\begin{translationbox}{Conditional Flag Influence in \Mathematica}
\begin{mycode}[language=myMMA, mathescape=true]
In[1] := Divide[1,Pi]*Integrate[Cos[z*Sin[$\theta$]], {$\theta$, 0, Pi},GenerateConditions -> False]
Out[1] = BesselJ[0, z]

In[2] := Divide[1,Pi]*Integrate[Cos[z*Sin[$\theta$]], {$\theta$, 0, Pi}]
Out[2] = BesselJ[0, Abs[z]] if $z \in \mathbb{R}$
\end{mycode}
\end{translationbox}
While confusing at the first glance, the output is not particularly wrong.
Since $J_0(z)$ is even in the second argument and along the Real line, the absolute value is simply redundant.

In case of~\dlmf{8.4.4}
\begin{equation}\label{eq:gamma-8-8-4}
\mathrm{\Gamma}(0,z) = \int_z^{\infty} t^{-1} e^{-t} \mathrm{d}t = E_1(z),
\end{equation}
we have not received any feedback from the developers.
For the difference between left- and right-hand side of the first equation
\begin{equation}
\mathrm{\Gamma}(0,z)\ \textcolor{Red}{-}\ \int_z^{\infty} t^{-1} e^{-t} \mathrm{d}t,
\end{equation}
\Mathematica{} conditionally returns $0$ if $\Re(z) > 0$ and $\Im(z) = 0$. 
We would expect $0$ without conditions.
Setting the problematic \verb|GenerateConditions| to \verb|False| returns $-\ln(z)$.
\begin{translationbox}{Second Example of Conditional Flag Influence in \Mathematica}
\begin{mycode}[language=myMMA, mathescape=true]
In[1] := Gamma[0,z] - Integrate[(t)^(-1)*Exp[-t], {t,z,Infinity}]
Out[1] = 0 if Re[z] > 0 && Im[z] == 0

In[2] := Gamma[0,z] - Integrate[(t)^(-1)*Exp[-t], {t,z,Infinity}, GenerateConditions -> False]
Out[2] = -Log[z]
\end{mycode}
\end{translationbox}
We noticed that another initial computation hook on the input could cause the issue.
For example, if we prevent instant evaluations on the input via \verb|Hold| and evaluate the expression on $z = \mathrm{i}$, \Mathematica{} returns $0. + 0.\mathrm{i}$.
Without \verb|Hold|, an evaluation on the same value returns \verb|undefined|.
\begin{translationbox}{Hold Inputs in \Mathematica}
\begin{mycode}[language=myMMA, mathescape=true]
In[1] := expr = Gamma[0,z] - Integrate[(t)^(-1)*Exp[-t], {t,z,Infinity}]
Out[1] = 0 if Re[z] > 0 && Im[z] == 0

In[2] := N[ReplaceAll[expr, {z -> I}]]
Out[2] = Undefined

In[3] := expr = Hold[Gamma[0,z] - Integrate[(t)^(-1)*Exp[-t], {t,z,Infinity}]]
Out[3] = Hold[Gamma[0,z] - Integrate[(t)^(-1)*Exp[-t], {t,z,Infinity}]]

In[4] := N[ReleaseHold[ReplaceAll[expr, {z -> I}]]]
Out[4] = 0. + 0. i
\end{mycode}
\end{translationbox}

All cases were handled by the \Mathematica{} team with the case ID 4664157.

\section{Evaluation Tables}\label{app:tables}

In this section, we provide three additional tables for our evaluation and translation results.
Table~\ref{tab:sample-evaluations}, provides three examples of our evaluations on the DLMF with different degrees of complexity. The first entry~\cite[\href{https://dlmf.nist.gov/1.4.E8}{(1.4.8)}]{DLMF}, for example, illustrates the difficulty of translating formulae from \LaTeX{} to CAS syntaxes even on a semantically enriched dataset like the DLMF.
Often, the arguments of a function in derivative notations are omitted since they can be deduced from the variable of differentiation. 
For example, $\tfrac{\mathrm{d}^2\;f}{\mathrm{d}x^2}$ the argument of the function $f(x)$ is omitted.
However, in this case \LaCASt{} is unable to correctly interpret $f$ as a function and presumed it to be a variable.
Unfortunately, not only caused this error a wrong translation but also produced a false positive evaluation because the symbolic simplification returned $0 = 0$ for
\begin{equation}
\verb|D[f, {x, 2}] == D[D[f, x], x]|.
\end{equation}
The other two examples in Table~\ref{tab:sample-evaluations}, even though more complex, illustrate the capability of \LaCASt{} and our evaluation pipeline.

Additionally, Table~\ref{tab:translations-table} and~\ref{tab:verification-table} show the number of translated and evaluated expressions for each chapter of the DLMF. 
For reference, Table~\ref{tab:dlmf} shows the full name of each chapter and the total number of displayed formulae according to the released dataset by Youssef and Miller~\cite{YoussefM20}. The actual number of functions may vary compared to Table~\ref{tab:translations-table}, because Youssef and Miller did not split multi-equations, and $\pm$ or $\mp$.
Hence, our final dataset consists of $10,930$ formulae.
Additionally, as described in our paper, we filter out non-semantic expressions, non-semantic macro definitions, ellipsis, approximations, and asymptotics. 
We ended up with $6,623$ test cases.
A more comprehensive table and all data is available at \url{https://lacast.wmflabs.org}.
For overview reasons, Table~\ref{tab:verification-table} only shows the results for translations to \Mathematica. For \Maple, see our website.

\newsavebox\translationOne
\begin{lrbox}{\translationOne}
 \begin{minipage}[t]{0.38\textwidth}
  \lstinline[language={[latex]TeX},mathescape,breaklines=true,basicstyle=\footnotesize\ttfamily]"D[f, {x, 2}] == D[D[f, x], x]"
 \end{minipage}
\end{lrbox}

\newsavebox\translationOneFixed
\begin{lrbox}{\translationOneFixed}
 \begin{minipage}[t]{0.48\textwidth}
  \lstinline[language={[latex]TeX},mathescape,breaklines=true,basicstyle=\footnotesize\ttfamily]"D[f[x], {x, 2}] == D[D[f[x], x], x]"
 \end{minipage}
\end{lrbox}

\newsavebox\translationTwoFirst
\begin{lrbox}{\translationTwoFirst}
 \begin{minipage}[t]{0.77\textwidth}
  \noindent\begin{lstlisting}[language={[latex]TeX},mathescape,xleftmargin=0pt,xrightmargin=0pt,breaklines=true,belowskip=0em,aboveskip=3pt,framesep=0pt,columns=fullflexible,basicstyle=\footnotesize\ttfamily,frame=tlbr,framerule=0pt,breakindent=0pt,numbers=none,numbersep=0pt,resetmargins=true,lineskip=3pt]
StruveH[\[Nu],z]-BesselY[\[Nu],z]==Divide[2*(Divide[1,2]*z)^\[Nu],Sqrt[Pi]*Gamma[\[Nu]+Divide[1,2]]]*Integrate[Exp[-z*t]*(1+(t)^(2))^(\[Nu]-Divide[1,2]),{t,0,Infinity},GenerateConditions->None]
  \end{lstlisting}
 \end{minipage}
\end{lrbox}

\newsavebox\translationThreeFirst
\begin{lrbox}{\translationThreeFirst}
 \begin{minipage}[t]{0.76\textwidth}
  \noindent\begin{lstlisting}[language={[latex]TeX},mathescape,xleftmargin=0pt,xrightmargin=0pt,breaklines=true,belowskip=0em,aboveskip=3pt,framesep=0pt,columns=fullflexible,basicstyle=\footnotesize\ttfamily,frame=tlbr,framerule=0pt,breakindent=0pt,numbers=none,numbersep=0pt,resetmargins=true,lineskip=3pt]
JacobiP[n,\[Alpha],\[Beta],x]==(2)^(-n)*Sum[Binomial[n+\[Alpha],\[ScriptL]]*Binomial[n+\[Beta],n-\[ScriptL]]*(x-1)^(n-\[ScriptL])*(x+1)^\[ScriptL],{\[ScriptL],0,n},GenerateConditions->None]
  \end{lstlisting}
 \end{minipage}
\end{lrbox}

\begin{table}[ht]
\vspace*{-1cm}
\centering
\caption{The table shows three sample cases of our evaluation pipeline from the DLMF. The translation shows the performed translations to Mathematica. The numeric column contains the number of successfully computed test cases. The constraints column contains all applied constraints including global constraints from Figure~\ref{fig:test-values}.}
\label{tab:sample-evaluations}
\begin{tabular}{c:c:c:c:Sc}
\hline\hline
\multicolumn{1}{l:}{\cite[\href{https://dlmf.nist.gov/1.4.E8}{(1.4.8)}]{DLMF}} & \multicolumn{4}{Sl}{$\frac{\mathrm{d}^2f}{\mathrm{d}x^2} = \frac{\mathrm{d}}{\mathrm{d}x}\left( \frac{\mathrm{d}f}{\mathrm{d}x}\right)$} \\\hdashline
\multicolumn{1}{Sl:}{\makecell[c]{Translation\\\wrong}} & \multicolumn{4}{l}{\makecell[l]{\usebox\translationOne \footnotesize(A correct translation requires the\\ \footnotesize argument for $f$, such as \usebox\translationOneFixed.)}} \\\hline
Symbolic & Numeric & Variables & Constraints & Test Values \\\hdashline
\correct & \correct (30/30) & $f$\ \wrong $,\ x$\ \correct & \makecell[c]{$x \in \mathbb{R},$\\ $\Re (x) > 0$} & \thead{$\scriptstyle x \in \left\{\tfrac{1}{2},\tfrac{3}{2},2\right\},$\\ $\scriptstyle f\in \left\{\pm\tfrac{1}{2},\pm\tfrac{3}{2},\pm 2, e^{2i\pi/3},\right.$\\ $\scriptstyle \left.e^{i\pi/6}, e^{-i\pi/3},e^{-5i\pi/6}\right\}$} \\\hline\hline
\multicolumn{1}{Sl:}{\cite[\href{https://dlmf.nist.gov/11.5.E2}{(11.5.2)}]{DLMF}} & \multicolumn{4}{Sl}{$\mathbf{K}_{\nu}(z) = \tfrac{2(\tfrac{1}{2}z)^{\nu}}{\sqrt{\pi}\Gamma(\nu+\tfrac{1}{2})}\int_{0}^{\infty}e^{-zt}(1+t^{2})^{\nu-\frac{1}{2}}\mathrm{d}t$} \\\hdashline
\multicolumn{1}{Sl:}{\makecell[c]{Translation\\\correct}} & \multicolumn{4}{l}{\thead{\usebox\translationTwoFirst}} \\\hline
Symbolic & Numeric & Variables & Constraints & Test Values \\\hdashline
\correct & \wrong (10/25) & $\nu, z$\ \correct & \makecell[c]{\footnotesize$-\pi < \operatorname{ph}(z) < \pi,$\\ \footnotesize$\Re (z) > 0,$\\ \footnotesize$\Re(\nu+\tfrac{1}{2}) > 0,$\\ \footnotesize$\Re(\nu+k+1) > 0,$\\ \footnotesize$\Re(-\nu+k+1) > 0,$\\ \footnotesize$\Re(n+\nu+\tfrac{3}{2}) > 0$} & \makecell[c]{$\nu,z \in \left\{\tfrac{1}{2},\tfrac{3}{2},2,\right.$\\ $\left.e^{i\pi/6},e^{-i\pi/3}\right\}$} \\\hline\hline
\multicolumn{1}{Sl:}{\cite[\href{https://dlmf.nist.gov/18.5.E8}{(18.5.8)}]{DLMF}} & \multicolumn{4}{Sl}{$P_n^{(\alpha,\beta)}(x) = 2^{-n}\sum_{\ell=0}^{n}\binom{n+\alpha}{\ell}\binom{n+\beta}{n-\ell}(x-1)^{n-\ell}(x+1)^{\ell}$} \\\hdashline
\multicolumn{1}{Sl:}{\makecell[c]{Translation\\\correct}} & \multicolumn{4}{l}{\thead{\usebox\translationThreeFirst}} \\\hline
Symbolic & Numeric & Variables & Constraints & Test Values \\\hdashline
\wrong & \correct (81/81) & $n, \alpha, \beta, x$\ \correct & \makecell[c]{$n\in\{1,2,3\},$\\ $x,\alpha,\beta \in \mathbb{R},$\\ $x,\alpha,\beta > 0$} & \makecell[c]{$n\in\{1,2,3\},$\\$\alpha,\beta,x \in \left\{\tfrac{1}{2},\tfrac{3}{2},2\right\}$} \\\hline\hline
\end{tabular}
\vspace*{-1cm}
\end{table}

{\small
\renewcommand{\arraystretch}{1.05}
\setlength\tabcolsep{6pt} 
\centering
\begin{table}[p]
\centering
\caption[Summary of the DLMF chapters.]{Summary of DLMF chapters with corresponding 2-letter codes (2C), chapter numbers (C{\tt\#}), 
chapter names of the DLMF chapters, and the total number of displayed formulas (i.e., without inline formulae) per chapter according to~\cite{YoussefM20} (F).}
\label{tab:dlmf}
\centering
\begin{tabular}[t]{ l:c:l:r }
\hline
{\bf 2C} & {\bf C\verb|#|} & \multicolumn{1}{c:}{\bf Chapter Name} & \multicolumn{1}{c}{\bf F} \\\hline
\rowcolor{TableRowColor}\verb|AL| & 1 & Algebraic and Analytic Methods & 571 \\
\verb|AS| & 2 & Asymptotic Approximations & 349 \\
\rowcolor{TableRowColor}\verb|NM| & 3 & Numerical Methods & 296 \\
\verb|EF| & 4 & Elementary Functions & 513 \\
\rowcolor{TableRowColor}\verb|GA| & 5 & Gamma Function & 161 \\
\verb|EX| & 6 & Exponential, Logarithmic, Sine, and Cosine Integrals & 100 \\
\rowcolor{TableRowColor}\verb|ER| & 7 & Error Functions, Dawson's and Fresnel Integrals & 137 \\
\verb|IG| & 8 & Incomplete Gamma and Related Functions & 240 \\
\rowcolor{TableRowColor}\verb|AI| & 9 & Airy and Related Functions & 230 \\
\verb|BS| & 10 & Bessel Functions & 696 \\
\rowcolor{TableRowColor}\verb|ST| & 11 & Struve and Related Functions & 149 \\
\verb|PC| & 12 & Parabolic Cylinder Functions & 177 \\
\rowcolor{TableRowColor}\verb|CH| & 13 & Confluent Hypergeometric Functions  & 372 \\
\verb|LE| & 14 & Legendre and Related Functions & 283 \\
\rowcolor{TableRowColor}\verb|HY| & 15 & Hypergeometric Function & 198 \\
\verb|GH| & 16 & Generalized Hypergeometric Functions \& Meijer $G$-Function & 99 \\
\rowcolor{TableRowColor}\verb|QH| & 17 & $q$-Hypergeometric and Related Functions & 181 \\
\verb|OP| & 18 & Orthogonal Polynomials & 502 \\
\rowcolor{TableRowColor}\verb|EL| & 19 & Elliptic Integrals & 464 \\
\verb|TH| & 20 & Theta Functions & 113 \\
\rowcolor{TableRowColor}\verb|MT| & 21 & Multidimensional Theta Functions & 59 \\
\verb|JA| & 22 & Jacobian Elliptic Functions & 258 \\
\rowcolor{TableRowColor}\verb|WE| & 23 & Weierstrass Elliptic and Modular Functions & 167 \\
\verb|BP| & 24 & Bernoulli and Euler Polynomials & 188 \\
\rowcolor{TableRowColor}\verb|ZE| & 25 & Zeta and Related Functions & 171 \\
\verb|CM| & 26 & Combinatorial Analysis & 201 \\
\rowcolor{TableRowColor}\verb|NT| & 27 & Functions of Number Theory & 132 \\
\verb|MA| & 28 & Mathieu Functions and Hill's Equation & 353 \\
\rowcolor{TableRowColor}\verb|LA| & 29 & Lam\'{e} Functions & 205 \\
\verb|SW| & 30 & Spheroidal Wave Functions & 114 \\
\rowcolor{TableRowColor}\verb|HE| & 31 & Heun Functions & 120 \\
\verb|PT| & 32 & Painlev\'{e} Transcendents & 286 \\
\rowcolor{TableRowColor}\verb|CW| & 33 & Coulomb Functions & 146 \\
\verb|TJ| & 34 & $3j$, $6j$, $9j$ Symbols & 64 \\
\rowcolor{TableRowColor}\verb|FM| & 35 & Functions of Matrix Argument & 62 \\
\verb|IC| & 36 & Integrals with Coalescing Saddles & 137 \\\hline
\multicolumn{3}{r}{$\Sigma$} & 8,494 \\
\end{tabular}
\end{table}
}

{\small
\renewcommand{\arraystretch}{1.06}
\setlength\tabcolsep{3.3pt} 
\centering
\begin{table}[p]
\centering
\caption[Overview of translations for DLMF chapters.]{Overview of translations for DLMF chapters. Table headings are 2C:~2-letter chapter codes;
C{\tt\#}:~chapter numbers;
F:~number of formulae;
T\textsubscript{old}:~number of translated expressions using old translator;
T\textsubscript{Map}, T\textsubscript{Math}:~number of translations with improved 
translator---Map for Maple and Math for Mathematica;
M\textsubscript{Map}, M\textsubscript{Math}:~number of failed translations due to missing 
macro translation; E:~number of other errors in the translation process. 
Best five performances are colored. Chapter codes are linked with our result page \demolink. 
}
\label{tab:translations-table}
\centering
\begin{tabular}[t]{ c : c : r | r r : r r : r r | r : r | r }
\hline
 {\bf 2C} & {\bf C\verb|#|} & \multicolumn{1}{c|}{\bf F} & \multicolumn{2}{c:}{\bf T\textsubscript{old}} & \multicolumn{2}{c:}{\bf T\textsubscript{Map}} & \multicolumn{2}{c|}{\bf T\textsubscript{Math}} & {\bf M\textsubscript{Map}} & {\bf M\textsubscript{Math}} & \multicolumn{1}{c}{\bf E} \\\hline
\rowcolor{TableRowColor}\lacastLink{Results_of_Algebraic_and_Analytic_Methods}{AL} &  1 & 227 &  60 & (26.4\%) & 102 & (44.9\%) & 103 & (45.4\%) &  79 &  78 & 46 \\
\lacastLink{Results_of_Asymptotic_Approximations}{AS} &  2 & 136 &  33 & (24.3\%) &  65 & (47.8\%) &  65 & (47.8\%) &  51 &  51 & 20 \\
\rowcolor{TableRowColor}\lacastLink{Results_of_Numerical_Methods}{NM} &  3 &  53 &  \cellcolor{CorrectBackgroundDLMFColor!30}36 & \cellcolor{CorrectBackgroundDLMFColor!30}(67.9\%) &  40 & (75.5\%) &  40 & (75.5\%) &   8 &   8 &  5 \\
\lacastLink{Results_of_Elementary_Functions}{EF} &  4 & 569 & 353 & (62.0\%) & 494 & (89.3\%) & \cellcolor{CorrectBackgroundDLMFColor}564 & \cellcolor{CorrectBackgroundDLMFColor}(99.1\%) &  58 &   4 &  1 \\
\rowcolor{TableRowColor}\lacastLink{Results_of_Gamma_Function}{GA} &  5 & 144 &  38 & (26.4\%) & \cellcolor{CorrectBackgroundDLMFColor!20}130 & \cellcolor{CorrectBackgroundDLMFColor!20}(93.5\%) & \cellcolor{CorrectBackgroundDLMFColor!50}139 & \cellcolor{CorrectBackgroundDLMFColor!50}(96.5\%) &   7 &   3 &  2 \\
\lacastLink{Results_of_Exponential,_Logarithmic,_Sine,_and_Cosine_Integrals}{EX} &  6 & 107 &  21 & (19.6\%) &  56 & (52.3\%) &  77 & (72.0\%) &  50 &  29 &  1 \\
\rowcolor{TableRowColor}\lacastLink{Results_of_Error_Functions,_Dawson’s_and_Fresnel_Integrals}{ER} &  7 & 149 &  35 & (23.5\%) & 101 & (67.8\%) & 120 & (80.5\%) &  47 &  28 &  1 \\
\lacastLink{Results_of_Incomplete_Gamma_and_Related_Functions}{IG} &  8 & 204 &  84 & (41.2\%) & 160 & (78.4\%) & 163 & (79.9\%) &  39 &  36 &  5 \\
\rowcolor{TableRowColor}\lacastLink{Results_of_Airy_and_Related_Functions}{AI} &  9 & 235 &  36 & (15.3\%) & 180 & (76.6\%) & 179 & (76.2\%) &  46 &  47 &  9 \\
\lacastLink{Results_of_Bessel_Functions}{BS} & 10 & 653 & 143 & (21.9\%) & 392 & (60.0\%) & 486 & (74.4\%) & 243 & 135 & 32 \\
\rowcolor{TableRowColor}\lacastLink{Results_of_Struve_and_Related_Functions}{ST} & 11 & 124 &  48 & (38.7\%) & \cellcolor{CorrectBackgroundDLMFColor}121 & \cellcolor{CorrectBackgroundDLMFColor}(97.6\%) & 112 & (90.3\%) &   3 &  12 &  0 \\
\lacastLink{Results_of_Parabolic_Cylinder_Functions}{PC} & 12 & 106 &  33 & (31.1\%) &  79 & (74.5\%) &  90 & (84.9\%) &  23 &   9 &  7 \\
\rowcolor{TableRowColor}\lacastLink{Results_of_Confluent_Hypergeometric_Functions}{CH} & 13 & 260 & 126 & (48.5\%) & \cellcolor{CorrectBackgroundDLMFColor!75}252 & \cellcolor{CorrectBackgroundDLMFColor!75}(96.9\%) & \cellcolor{CorrectBackgroundDLMFColor!75}254 & \cellcolor{CorrectBackgroundDLMFColor!75}(97.7\%) &   3 &   1 &  5 \\
\lacastLink{Results_of_Legendre_and_Related_Functions}{LE} & 14 & 238 & \cellcolor{CorrectBackgroundDLMFColor!50}166 & \cellcolor{CorrectBackgroundDLMFColor!50}(69.7\%) & \cellcolor{CorrectBackgroundDLMFColor!50}230 & \cellcolor{CorrectBackgroundDLMFColor!50}(96.6\%) & \cellcolor{CorrectBackgroundDLMFColor!30}229 & \cellcolor{CorrectBackgroundDLMFColor!30}(96.2\%) &   5 &   6 &  3 \\
\rowcolor{TableRowColor}\lacastLink{Results_of_Hypergeometric_Function}{HY} & 15 & 206 & \cellcolor{CorrectBackgroundDLMFColor!75}148 & \cellcolor{CorrectBackgroundDLMFColor!75}(71.8\%) & \cellcolor{CorrectBackgroundDLMFColor!30}198 & \cellcolor{CorrectBackgroundDLMFColor!30}(96.1\%) & \cellcolor{CorrectBackgroundDLMFColor!20}197 & \cellcolor{CorrectBackgroundDLMFColor!20}(95.6\%) &   3 &   4 &  5 \\
\lacastLink{Results_of_Generalized_Hypergeometric_Functions_and_Meijer_G-Function}{GH} & 16 &  53 &  20 & (37.7\%) &  23 & (43.4\%) &  25 & (47.2\%) &  27 &   1 & 27 \\
\rowcolor{TableRowColor}\lacastLink{Results_of_q-Hypergeometric_and_Related_Functions}{QH} & 17 & 175 &   1 & ( 0.6\%) &  53 & (30.3\%) & 124 & (70.8\%) & 112 &  35 & 16 \\
\lacastLink{Results_of_Orthogonal_Polynomials}{OP} & 18 & 468 & 132 & (28.2\%) & 235 & (50.2\%) & 288 & (61.5\%) & 203 & 149 & 31 \\
\rowcolor{TableRowColor}\lacastLink{Results_of_Elliptic_Integrals}{EL} & 19 & 516 & 103 & (20.0\%) & 252 & (48.8\%) & 416 & (80.6\%) & 250 &  84 & 16 \\
\lacastLink{Results_of_Theta_Functions}{TH} & 20 & 128 &  52 & (40.6\%) &  98 & (76.6\%) &  98 & (76.6\%) &   8 &   8 & 22 \\
\rowcolor{TableRowColor}\lacastLink{Results_of_Multidimensional_Theta_Functions}{MT} & 21 &  32 &   0 & ( 0.0\%) &   0 & ( 0.0\%) &   0 & ( 0.0\%) &  30 &  30 &  2 \\
\lacastLink{Results_of_Jacobian_Elliptic_Functions}{JA} & 22 & 264 & 115 & (43.6\%) & 232 & (87.9\%) & 238 & (90.2\%) &  27 &  21 &  5 \\
\rowcolor{TableRowColor}\lacastLink{Results_of_Weierstrass_Elliptic_and_Modular_Functions}{WE} & 23 & 164 &   7 & ( 4.3\%) &  19 & (11.6\%) &  34 & (20.7\%) & 125 & 112 & 18 \\
\lacastLink{Results_of_Bernoulli_and_Euler_Polynomials}{BP} & 24 & 175 &  31 & (17.7\%) & 117 & (67.2\%) & 148 & (84.6\%) &  35 &  15 & 12 \\
\rowcolor{TableRowColor}\lacastLink{Results_of_Zeta_and_Related_Functions}{ZE} & 25 & 154 &  28 & (18.2\%) & 124 & (80.5\%) & 120 & (77.9\%) &  26 &  30 &  4 \\
\lacastLink{Results_of_Combinatorial_Analysis}{CM} & 26 & 136 &  31 & (22.8\%) &  78 & (57.3\%) &  87 & (64.0\%) &  54 &  42 &  7 \\
\rowcolor{TableRowColor}\lacastLink{Results_of_Functions_of_Number_Theory}{NT} & 27 &  79 &   5 & ( 6.3\%) &  26 & (32.9\%) &  15 & (19.0\%) &  38 &  49 & 15 \\
\lacastLink{Results_of_Mathieu_Functions_and_Hill’s_Equation}{MA} & 28 & 267 &  52 & (19.5\%) &  97 & (36.3\%) & 110 & (41.2\%) & 138 & 125 & 32 \\
\rowcolor{TableRowColor}\lacastLink{Results_of_Lam\%C3\%A9_Functions}{LA} & 29 & 111 &  11 & ( 9.9\%) &  23 & (20.7\%) &  22 & (19.8\%) &  79 &  80 &  9 \\
\lacastLink{Results_of_Spheroidal_Wave_Functions}{SW} & 30 &  71 &  14 & (19.7\%) &  19 & (26.8\%) &  26 & (36.6\%) &  47 &  39 &  6 \\
\rowcolor{TableRowColor}\lacastLink{Results_of_Heun_Functions}{HE} & 31 &  35 &  \cellcolor{CorrectBackgroundDLMFColor}29 & \cellcolor{CorrectBackgroundDLMFColor}(82.8\%) &  22 & (62.8\%) &  15 & (42.8\%) &   9 &  16 &  4 \\
\lacastLink{Results_of_Painlev\%C3\%A9_Transcendents}{PT} & 32 &  67 &  \cellcolor{CorrectBackgroundDLMFColor!20}43 & \cellcolor{CorrectBackgroundDLMFColor!20}(64.2\%) &  57 & (85.1\%) &  57 & (85.1\%) &   0 &   0 & 10 \\
\rowcolor{TableRowColor}\lacastLink{Results_of_Coulomb_Functions}{CW} & 33 & 108 &  21 & (19.4\%) &  14 & (13.0\%) &  11 & (10.2\%) &  80 &  86 & 11 \\
\lacastLink{Results_of_3j,6j,9j_Symbols}{TJ} & 34 &  57 &   0 & ( 0.0\%) &   1 & ( 1.8\%) &  37 & (64.9\%) &  46 &   4 & 16 \\
\rowcolor{TableRowColor}\lacastLink{Results_of_Functions_of_Matrix_Argument}{FM} & 35 &  46 &   0 & ( 0.0\%) &   0 & ( 0.0\%) &   0 & ( 0.0\%) &  36 &  36 & 10 \\
\lacastLink{Results_of_Integrals_with_Coalescing_Saddles}{IC} & 36 & 106 &  12 & (11.3\%) &  24 & (22.6\%) &  24 & (22.6\%) &  79 &  79 &  3 \\
\hline\hline
\multicolumn{2}{c:}{$\Sigma$} & 6,623 & 2,067 & (31.2\%) & 4,114 & (62.1\%) & 4,713 & (71.2\%) & 2,114 & 1,492 & 418 \\\hline
\end{tabular}
\end{table}
}

{\small
\renewcommand{\arraystretch}{1.06}
\setlength\tabcolsep{3.4pt} 
\begin{table}[p]
\centering
\caption[Overview of symbolic and numeric evaluations for DLMF chapters.]{Overview of symbolic and numeric evaluations for DLMF chapters as in Table~\ref{tab:translations-table}. 
Table headings are T\textsubscript{Math}:~number of successfully translations to Mathetmatica;
S\textsubscript{success}, S\textsubscript{fail}:~number of successful and failed symbolic verifications 
(for translated expressions only) respectively;
N\textsubscript{success}, N\textsubscript{fail}:~number of successful and remaining failed numeric 
(for failed symbolical tests only) respectively.
P, T:~number of partial (at least one test was successful) and total failed numeric tests.
Best five performances are colored. Chapter codes are linked with our result page \demolink.}
\label{tab:verification-table}
\centering
\begin{tabular}[t]{ c : c : r | r r : r | r r : r r : r : r }
\hline
{\bf 2C} & {\bf C\verb|#|} & \multicolumn{1}{c|}{\bf T\textsubscript{Math}} & 
\multicolumn{2}{c:}{\bf S\textsubscript{success}} & \multicolumn{1}{c|}{\bf S\textsubscript{fail}} & 
\multicolumn{2}{c:}{\bf N\textsubscript{success}} & \multicolumn{1}{c}{\bf N\textsubscript{fail}} & \multicolumn{1}{r:}{\bf [ P / T ]} & {\bf A} & {\bf E} \\\hline
\rowcolor{TableRowColor}\lacastLink{Results_of_Algebraic_and_Analytic_Methods}{AL} &  1 & 103 &  34 & (33.0\%) &  69 &  14 & (20.3\%) &  40 & [ \phantom{0}9 /  31 ] &  11 &   4 \\
\lacastLink{Results_of_Asymptotic_Approximations}{AS} &  2 &  65 &   6 & ( 9.2\%) &  59 &   4 & ( 6.8\%) &  38 & [ \phantom{0}6 /  32 ] &   7 &   9 \\
\rowcolor{TableRowColor}\lacastLink{Results_of_Numerical_Methods}{NM} &  3 &  40 &   5 & (12.5\%) &  35 &   0 & ( 0.0\%) &  29 & [ \phantom{0}8 /  21 ] &   6 &   0 \\
\lacastLink{Results_of_Elementary_Functions}{EF} &  4 & 564 & \cellcolor{CorrectBackgroundDLMFColor}304 & \cellcolor{CorrectBackgroundDLMFColor}(53.9\%) & 260 & \cellcolor{CorrectBackgroundDLMFColor!50}110 & \cellcolor{CorrectBackgroundDLMFColor!50}(42.3\%) & 146 & [ 55 /  91 ] &   2 &   0 \\
\rowcolor{TableRowColor}\lacastLink{Results_of_Gamma_Function}{GA} &  5 & 139 &  \cellcolor{CorrectBackgroundDLMFColor!75}65 & \cellcolor{CorrectBackgroundDLMFColor!75}(46.8\%) &  74 &  \cellcolor{CorrectBackgroundDLMFColor!30}30 & \cellcolor{CorrectBackgroundDLMFColor!30}(40.5\%) &  20 & [ \phantom{0}9 /  11 ] &  13 &   9 \\
\lacastLink{Results_of_Exponential,_Logarithmic,_Sine,_and_Cosine_Integrals}{EX} &  6 &  77 &  18 & (23.4\%) &  59 &  \cellcolor{CorrectBackgroundDLMFColor!20}23 & \cellcolor{CorrectBackgroundDLMFColor!20}(39.0\%) &  32 & [ \phantom{0}6 /  26 ] &   4 &   0 \\
\rowcolor{TableRowColor}\lacastLink{Results_of_Error_Functions,_Dawson’s_and_Fresnel_Integrals}{ER} &  7 & 120 &  45 & (37.5\%) &  75 &  21 & (28.0\%) &  43 & [ 13 /  30 ] &   9 &   1 \\
\lacastLink{Results_of_Incomplete_Gamma_and_Related_Functions}{IG} &  8 & 163 & \cellcolor{CorrectBackgroundDLMFColor!20} 65 & \cellcolor{CorrectBackgroundDLMFColor!20}(39.9\%) &  98 &  22 & (22.4\%) &  44 & [ 19 /  25 ] &  16 &  15 \\
\rowcolor{TableRowColor}\lacastLink{Results_of_Airy_and_Related_Functions}{AI} &  9 & 179 &  69 & (38.5\%) & 110 &  30 & (27.3\%) &  58 & [ 38 /  20 ] &  14 &   7 \\
\lacastLink{Results_of_Bessel_Functions}{BS} & 10 & 486 & 115 & (23.7\%) & 371 &  90 & (24.2\%) & 151 & [ 57 /  94 ] &  92 &  18 \\
\rowcolor{TableRowColor}\lacastLink{Results_of_Struve_and_Related_Functions}{ST} & 11 & 112 &  36 & (32.1\%) &  76 &  21 & (27.6\%) &  33 & [ \phantom{0}8 /  25 ] &  10 &  11 \\
\lacastLink{Results_of_Parabolic_Cylinder_Functions}{PC} & 12 &  90 &  18 & (20.0\%) &  72 &  13 & (18.0\%) &  43 & [ 15 /  28 ] &  12 &   3 \\
\rowcolor{TableRowColor}\lacastLink{Results_of_Confluent_Hypergeometric_Functions}{CH} & 13 & 254 &  69 & (27.2\%) & 185 &  23 & (12.4\%) &  95 & [ 59 /  36 ] &  45 &  21 \\
\lacastLink{Results_of_Legendre_and_Related_Functions}{LE} & 14 & 229 &  30 & (13.1\%) & 199 &  59 & (29.6\%) &  92 & [ 54 /  38 ] &  41 &   5 \\
\rowcolor{TableRowColor}\lacastLink{Results_of_Hypergeometric_Function}{HY} & 15 & 197 &  53 & (26.9\%) & 144 &  23 & (16.0\%) &  77 & [ 52 /  25 ] &  29 &   6 \\
\lacastLink{Results_of_Generalized_Hypergeometric_Functions_and_Meijer_G-Function}{GH} & 16 &  25 &   2 & ( 8.0\%) &  23 &   1 & ( 4.3\%) &  10 & [ \phantom{0}7 / \phantom{0}3 ] &   9 &   2 \\
\rowcolor{TableRowColor}\lacastLink{Results_of_q-Hypergeometric_and_Related_Functions}{QH} & 17 & 124 &   6 & ( 4.8\%) & 118 &  13 & (11.0\%) &  57 & [ 52 / \phantom{0}5 ] &  39 &   5 \\
\lacastLink{Results_of_Orthogonal_Polynomials}{OP} & 18 & 288 & 101 & (35.1\%) & 185 &  45 & (24.3\%) &  68 & [ 31 /  37 ] &  52 &  12 \\
\rowcolor{TableRowColor}\lacastLink{Results_of_Elliptic_Integrals}{EL} & 19 & 416 &  51 & (12.2\%) & 365 &  18 & ( 4.9\%) & 264 & [ 49 /215 ] &  61 &  15 \\
\lacastLink{Results_of_Theta_Functions}{TH} & 20 &  98 &   1 & ( 1.0\%) &  97 &  33 & (34.0\%) &  40 & [ 25 /  15 ] &  24 &   0 \\
\rowcolor{TableRowColor}\lacastLink{Results_of_Multidimensional_Theta_Functions}{MT} & 21 &   0 & \multicolumn{2}{c:}{-} & - & \multicolumn{2}{c:}{-} & \multicolumn{2}{c:}{-} & - & - \\
\lacastLink{Results_of_Jacobian_Elliptic_Functions}{JA} & 22 & 238 &  30 & (12.6\%) & 206 &  22 & (10.7\%) & 131 & [ 39 /  92 ] &  51 &   0 \\
\rowcolor{TableRowColor}\lacastLink{Results_of_Weierstrass_Elliptic_and_Modular_Functions}{WE} & 23 &  34 &   4 & (11.8\%) &  30 &   2 & ( 6.7\%) &  23 & [ \phantom{0}9 /  14 ] &   2 &   3 \\
\lacastLink{Results_of_Bernoulli_and_Euler_Polynomials}{BP} & 24 & 148 &  23 & (15.5\%) & 125 &  \cellcolor{CorrectBackgroundDLMFColor!75}78 & \cellcolor{CorrectBackgroundDLMFColor!75}(62.4\%) &  33 & [ 22 /  11 ] &  14 &   0 \\
\rowcolor{TableRowColor}\lacastLink{Results_of_Zeta_and_Related_Functions}{ZE} & 25 & 120 &  \cellcolor{CorrectBackgroundDLMFColor!50}48 & \cellcolor{CorrectBackgroundDLMFColor!50}(40.0\%) &  72 &  22 & (30.5\%) &  22 & [ \phantom{0}6 /  16 ] &  22 &   3 \\
\lacastLink{Results_of_Combinatorial_Analysis}{CM} & 26 &  87 &  19 & (21.8\%) &  68 &  \cellcolor{CorrectBackgroundDLMFColor}44 & \cellcolor{CorrectBackgroundDLMFColor}(64.7\%) &  18 & [ 10 / \phantom{0}8 ] &   5 &   1 \\
\rowcolor{TableRowColor}\lacastLink{Results_of_Functions_of_Number_Theory}{NT} & 27 &  15 &   \cellcolor{CorrectBackgroundDLMFColor!50}6 & \cellcolor{CorrectBackgroundDLMFColor!50}(40.0\%) &   9 &   3 & (33.3\%) &   6 & [ \phantom{0}3 / \phantom{0}3 ] &   0 &   0 \\
\lacastLink{Results_of_Mathieu_Functions_and_Hill’s_Equation}{MA} & 28 & 110 &   7 & ( 6.4\%) & 103 &   3 & ( 2.9\%) &  48 & [ 13 /  35 ] &  33 &  17 \\
\rowcolor{TableRowColor}\lacastLink{Results_of_Lam\%C3\%A9_Functions}{LA} & 29 &  22 &   0 & ( 0.0\%) &  22 &   0 & ( 0.0\%) &  21 & [ \phantom{0}1 /  20 ] &   0 &   1 \\
\lacastLink{Results_of_Spheroidal_Wave_Functions}{SW} & 30 &  26 &   0 & ( 0.0\%) &  26 &   0 & ( 0.0\%) &  19 & [ \phantom{0}2 /  17 ] &   5 &   1 \\
\rowcolor{TableRowColor}\lacastLink{Results_of_Heun_Functions}{HE} & 31 &  15 &   2 & (13.3\%) &  13 &   0 & ( 0.0\%) &   8 & [ \phantom{0}0 / \phantom{0}8 ] &   5 &   0 \\
\lacastLink{Results_of_Painlev\%C3\%A9_Transcendents}{PT} & 32 &  57 &   3 & ( 5.3\%) &  54 &   0 & ( 0.0\%) &  41 & [ \phantom{0}2 /  39 ] &   8 &   5 \\
\rowcolor{TableRowColor}\lacastLink{Results_of_Coulomb_Functions}{CW} & 33 &  11 &   0 & ( 0.0\%) &  11 &   0 & ( 0.0\%) &  11 & [ \phantom{0}2 / \phantom{0}9 ] &   0 &   0 \\
\lacastLink{Results_of_3j,6j,9j_Symbols}{TJ} & 34 &  37 &   0 & ( 0.0\%) &  37 &  14 & (37.8\%) &  10 & [ \phantom{0}5 / \phantom{0}5 ] &  13 &   0 \\
\rowcolor{TableRowColor}\lacastLink{Results_of_Functions_of_Matrix_Argument}{FM} & 35 &   0 & \multicolumn{2}{c:}{-} & - & \multicolumn{2}{c:}{-} & \multicolumn{2}{c:}{-} & - & - \\
\lacastLink{Results_of_Integrals_with_Coalescing_Saddles}{IC} & 36 &  24 &   0 & ( 0.0\%) &  24 &   3 & (12.5\%) &  13 & [ \phantom{0}1 /  12 ] &   1 &   6 \\
\hline\hline
\multicolumn{2}{c:}{$\Sigma$} & 4,713 & 1,235 & (26.2\%) & 3,474 &  784 & (22.6\%) & 1,784 & [687 / 1,097] &  655 &  180 \\ \hline
\end{tabular}
\end{table}
}

\end{document}
\endinput